\documentclass{lmcs}
\pdfoutput=1
\usepackage[utf8]{inputenc}

\usepackage{lastpage}
\lmcsdoi{21}{1}{26}
\lmcsheading{}{\pageref{LastPage}}{}{}%
{Jan.~26,~2024}{Mar.~18,~2025}{}

\usepackage[T1]{fontenc}

\usepackage{graphicx}
\usepackage{amsmath, amssymb}
\usepackage{tikz}
\usetikzlibrary{patterns, intersections, positioning, shapes.geometric, calc,
automata, arrows}
\usepackage[compatibility=false]{caption}
\usepackage{listings}
\usepackage{microtype}
\usepackage{wasysym}
\usepackage{multirow}
\usepackage{xcolor}
\usepackage{soul,cancel}
\usepackage{sidecap}
\usepackage{newtxmath}
\usepackage{sidecap}
\usepackage[normalem]{ulem}
\usepackage{thmtools}
\usepackage{hyperref}
\usepackage{marvosym}
\usepackage{subfig}
\usepackage{mathtools}

\definecolor{lightgray}{rgb}{0.75, 0.75, 0.75}

\newcommand{\Outc}{\textit{Outc}}
\newcommand{\winp}{{\textit{Win$\Pi$}}}
\newcommand{\wins}{{\textit{Win$\Sigma$}}}
\newcommand{\winrop}{\textit{WinReg}}
\newcommand{\pref}{{\textit{Pref}}}
\newcommand{\ind}{\textit{WinInd}}
\newcommand{\finally}{R}
\newcommand{\act}{\textrm{Act}}
\newcommand{\sigmabar}{\bar{\sigma}}
\newcommand{\moves}{\textit{Moves}}
\newcommand{\cost}{\costop}
\newcommand{\costop}{\textit{Cost}}

\newcommand{\playpref}{\Pi}
\newcommand{\pennygame}{\game_\text{\EUR}}

\newcommand{\game}{{G}}

\newcommand{\dattrop}{\textit{dJAttr}}
\newcommand{\dattrg}{\dattrop(\game,\finally)}
\newcommand{\dattrk}[1]{\dattrop^{#1}(\game,\finally)}

\newcommand{\cpre}{\textit{CPre}}
\newcommand{\pre}{\textit{Pre}}
\newcommand{\dpre}{\textit{dPre}}
\newcommand{\sattr}{\mathit{Safe}}
\newcommand{\rankop}{\textit{Rank}}
\newcommand{\arankop}{A\rankop}
\newcommand{\jrankop}{J\rankop}
\newcommand{\arankg}{\arankop}
\newcommand{\jrankg}{\jrankop}
\newcommand{\parankop}{pA\rankop}
\newcommand{\pjrankop}{pJ\rankop}
\newcommand{\djrankop}{dJ\rankop}
\newcommand{\parankg}{\parankop}
\newcommand{\pjrankg}{\pjrankop}
\newcommand{\djrankg}{\djrankop}

\newcommand{\attrf}{\attrop(\game,\finally)}

\newcommand{\attrfi}[1]{\attrop^{#1}(\game,\finally)}

\newcommand{\attrop}{\textit{Attr}}
\newcommand{\pattrf}{\pattrop_1(\game,\finally)}
\newcommand{\pattrfi}[1]{\pattrop^{#1}_1(\game,\finally)}

\newcommand{\pattrsi}[1]{\pattrop^{#1}_2(\game,\finally)}
\newcommand{\pattrop}{\textit{pAttr}}

\newcommand{\reachop}{\textit{Reach}}
\newcommand{\reach}{{\reachop(\game,\finally)}}

\newcommand{\jokerg}{\jokerop(\game,\finally)}
\newcommand{\pjokerg}{\pjokerop(\game,\finally)}
\newcommand{\joker}[1]{\jokerop^{#1}(\game,\finally)}
\newcommand{\pjoker}[1]{\pjokerop^{#1}(\game,\finally)}
\newcommand{\jokerreal}[1]{\jokerop^{#1}_\jokersymb(\game,\finally)}
\newcommand{\pjokerreal}[1]{\pjokerop^{#1}_\jokersymb(\game,\finally)}
\newcommand{\jokerop}{\textit{JAttr}}
\newcommand{\pjokerop}{p\jokerop}
\newcommand{\jstate}{\jokerop_\jokersymbreal(\game,\finally)}
\newcommand{\pjstate}{\pjokerop_\jokersymbreal(\game,\finally)}

\newcommand{\jokersymb}{\varspadesuit}
\newcommand{\jokersymbreal}{{\spadesuit}}
\newcommand{\jokergame}{{\game^\jokersymb}}

\newcommand{\Distr}{{\textit{Distr}}}

\newcommand{\gameaorb}{G_{a \wedge b}}
\newcommand{\gamedist}{\game_{dist}}

\newcommand{\prooflink}[1]{\noindent\hyperlink{#1}{\textit{See proof in appendix.}}}

\begin{document}

\title[Semi-cooperative games via Joker moves]{With a little help from your friends: semi-cooperative games via Joker moves}

\author[P.~van~den~Bos]{Petra {van den Bos}\lmcsorcid{0000-0002-9212-1525}}[a]
\author[M.~Stoelinga]{Marielle Stoelinga \lmcsorcid{0000-0001-6793-8165}}[a,b]

\address{Formal Methods \& Tools, University of Twente, the Netherlands}
\email{\{p.vandenbos, m.i.a.stoelinga\}@utwente.nl}

\address{Department of Software Science,
Radboud University, the Netherlands}

\begin{abstract}
This paper coins the notion of Joker games, a variant of concurrent games where the players are not strictly adversarial. Instead, Player 1 can get help from Player 2 by playing a Joker move. We formalize these games as cost games and develop strategies that minimize the use of Jokers -- viewed as costs -- to secure a win with the least possible help.
Our investigation studies the theoretical underpinnings of these games and their associated Joker strategies. In particular, when comparing our cost-minimal strategies with admissible strategies, we find out that they differ.
Moreover, while randomization can be beneficial in conventional concurrent games, it does not aid in winning Joker games, although it can help reduce the number of needed Jokers. We also enhance our framework by introducing a secondary objective, namely by minimizing the number of moves executed by a Joker strategy.
Finally, we demonstrate the practical advantages of our approach by applying it to test generation in model-based testing.
\end{abstract}

\maketitle

\section{Introduction}

\paragraph{Winning strategies.} We study 2 player concurrent games played on a game graph with reachability objectives, i.e., the goal for Player 1 is to reach a set of goal states $R$. A central notion in such games is the concept of a {\em winning strategy}, which assigns ---in states where this is possible
---moves to Player 1 so that she always reaches the goal $R$, no matter how Player 2 plays.

Concurrent reachability games have abundant applications, e.g. controller synthesis \cite{bertsekas}, 
assume/guarantee reasoning \cite{ChatterjeeH07-assume-guarantee}, interface theory \cite{DBLP:journals/entcs/AlfaroS04},
security \cite{conf-gamesec/2016} and test case optimization \cite{DBLP:conf/fortest/HesselLMNPS08,realtimegames}.
However, it is widely acknowledged 
\cite{Berwanger07,Faella09,ChatterjeeH07-assume-guarantee}
that the concept of a winning strategy is often too strong in these applications: unlike games like chess, the opponent does not try to make Player 1's life as hard as possible, but rather, the behaviour of Player 2 is unknown. 
Moreover, winning strategies do not prescribe any move in states where Player 1 cannot win for sure. This is a major drawback: even if Player 1 cannot win, some strategies are still better than others. Such better strategies select best-effort moves that maximize the possibility of winning in some next state, even though the game cannot be won for sure. 
Several concepts have been proposed to formalize best-effort strategies \cite{synthesisresilient,synthesissafety,BARTOCCI2021229,AnandMNS23}. This paper proposes a new approach, using \emph{Jokers}.

\paragraph{Our approach}
In this paper, we introduce the notion of {\em Joker games}, which  fruitfully builds upon techniques for robust strategy synthesis from Neider et al.  \cite{synthesisresilient,synthesissafety}. 
We define a \emph{Joker strategy} as a best-effort strategy in cases where Player 1 cannot win:  
Joker games allow Player 1, in addition to regular moves with her own actions, to play a so-called {\em Joker}. 
When Player 1 plays a Joker, she not only choose her own move, but also the opponent's move. In this way, Player 2 helps Player 1 in reaching the goal. 
Next, we minimize the number of Jokers needed to win the game, 
so that Player 1 only calls for help when needed, using an attractor construction.
The obtained Joker-minimal strategies are naturally formalized as cost-minimal strategies in a priced game.

We note that in applications, where it is not possible to actually play a Joker, \emph{Joker-inspired} strategies can be used. Such a strategy is defined the same as a Joker strategy, except that when the Joker strategy would they play a Joker action, the Joker-inspired strategy plays the Player 1 action of that Joker action. When a Player 1 action is played instead of a Joker, Player 1 can still hope that Player 2 chooses a cooperative action, e.g. the Player 2 action from the Joker action. By using a Joker-inspired strategy, Player 1 requires the minimum number of cooperative Player 2 actions for winning, thus minimizing the `luck' required for winning. This is fruitful in cases where the opponent is not strictly adversarial. Also we note that even when Player 2 plays a non-cooperative action, our strategy will steer back to the goal from the reached state, if winning is still possible.

Our Joker games are concurrent and nondeterministic. 
Concurrency allows us to faithfully model and handle strategy synthesis for applications with concurrent interactions, where waiting for your turn cannot be assumed.
Moreover, nondeterminism allows that outcome of the moves may lead to several next states. In this way, our games express that outcomes cannot be controlled. We do not assign probabilities to quantify nondeterminism, because especially for opponent actions, 
 the probabilities are often unknown for real applications.

Concurrency and nondeterminism are essential in several applications. In  game-based testing, where test cases are constructed from synthesized strategies, nondeterminism is needed to faithfully model the interaction between the tester and the system-under-test \cite{mbtgames}. Our games can also model other concurrent and distributed applications, e.g. where two services or agents interact with each other via non-synchronized communication channels, or where a plant operates in its (uncontrollable!) environment.

One of the contributions of this paper is that we present the link of our Joker strategies with cost-minimal strategies.
We show that cost-minimal strategies are more general than attractor strategies: all attractor strategies are cost-minimal, but not all cost-minimal strategies are attractor strategies. In particular, non-attractor strategies may outperform attractor strategies with respect to other objectives like the number of moves needed to reach the goal. In this paper, we study and define multi-objective strategies that minimize both the number of Jokers and moves.

Furthermore, we establish several new results for Joker games: (1) While concurrent games require randomized strategies to win a game, this is not true for the specific Joker moves: these can be played deterministically. (2) If we play a strategy with $n$ Jokers, each play contains exactly $n$ Joker moves. 
(3)
Even with deterministic strategies, Joker games are determined. (4) The classes of Joker strategies and admissible strategies do not coincide. 

As an extension to result (1) is that although randomization is not necessary to win in Joker games, randomization can be used to reduce the number of required Jokers for winning (with probability 1). Also, result (4) is extended by explaining the differences between Joker strategies and admissible strategies, with examples, and by showing that against a restricted opponent, without memory, Joker strategies and admissible strategies do coincide.

Finally, we illustrate how Joker strategies can be applied in practice, by extracting test cases from Joker-inspired strategies. Here we use techniques from previous work \cite{mbtgames}, to translate game strategies to test cases. In particular, a test case provides Player 1 actions as input to the System-Under-Test. We refer to \cite{mbtgames} for related work on game-based approaches for testing. The experiments of this paper on four classes of case studies show that obtained test cases outperform the standard testing approach of random testing. Specifically,
Joker-inspired test cases reach the goal more often than random ones, and require fewer steps to do so.

\paragraph{Contributions.} Summarizing, the main contributions of this paper are:
\begin{enumerate}
    \item We formalize the minimum help Player 1 needs from Player 2 to win as cost-minimal strategies in a Joker game.
    \item We establish several properties: the minimum number of Jokers equals minimum cost, each play of a Joker strategy uses $n$ Jokers, Joker game determinacy, Jokers can be played deterministically, in randomized setting, and admissible strategies do not coincide with cost-minimal strategies.
    \item We refine our Joker approach with second objective of the number of moves.
    \item We illustrate the benefits of our approach for test case generation.
\end{enumerate}

We note that, compared to conference paper \cite{ForteBosS23}, we include the proofs of our theorems (in the appendix), we provide a constructive definition for our multi-objective strategies, we provide a completed comparison of Joker strategies with randomized strategies (with all required definitions)
, and we solve the open conjecture of the conference paper \cite{ForteBosS23} on admissible strategies.

\paragraph{Paper organization.}
Section~\ref{sec:relwork} compares our work with related work.
Section~\ref{sec:games} recapitulates concurrent games.  Section~\ref{sec:joker-games} introduces Joker games, and Section~\ref{sec:properties-joker-games} investigates their properties.
In \autoref{sec:admissible} we study the relation between cost-minimal and admissible strategies, and in Sect~\ref{sec:short} we study multi-objective Joker strategies
Furthermore, we study randomization in \autoref{sec:randstrats}
Then, in Section ~\ref{sec:experiments} we apply Joker strategies to test case generation. Finally, Section~\ref{sec:concl} concludes the paper. 
Proofs of the theorems of this paper are given in the Appendix 
The artifact
of our experimental results of \autoref{sec:experiments} is provided in \cite{artefact}.

\section{Related work} \label{sec:relwork}
\paragraph{Dominance.}
Berwanger \cite{Berwanger07}  coins the notion of strategy dominance, where one strategy is better than another if it wins against more opponent strategies.
The maximal elements in the lattice of Player-1 strategies,  called {\em admissible} strategies, are proposed as best strategies in losing states.
We show in Section~\ref{sec:admissible} that Joker strategies correspond to admissible strategies, against a restricted opponent, without memory.

\paragraph{Attractor strategies.}
Dallal et al. consider `better' actions outside the winning set \cite{synthesissafety}, and resilience against disturbances of the opponent \cite{synthesisresilient}.
In addition, Bartocci et al. \cite{BARTOCCI2021229} perform numeric optimization on cooperative reachability games to obtain test cases.
Recently, Anand et al. \cite{AnandMNS23} introduced adequately permissive assumptions for strategy synthesis for $\omega$-regular games. 

While the construction for Joker strategies closely follows the (attractor) constructions given in \cite{synthesissafety,BARTOCCI2021229,AnandMNS23}, our paper provides different and additional results. First, the games in \cite{synthesisresilient,synthesissafety,BARTOCCI2021229,AnandMNS23} are turn-based and deterministic, whereas ours are concurrent and nondeterministic.

Secondly, while \cite{synthesissafety,BARTOCCI2021229,AnandMNS23} compute optimal strategies in the form of attractor strategies, the works \cite{synthesisresilient,synthesissafety,BARTOCCI2021229,AnandMNS23} do not present the link with cost-minimal strategies, like we do for our Joker strategies. 

Thirdly, existing works \cite{synthesisresilient,AnandMNS23} consider more advanced winning conditions for B\"uchi and parity games than reachability, but they do this for a single winning condition $\Phi$. We also study strategies for the multi-objective of minimizing both the number of Jokers and moves.

A fourth difference with existing works \cite{synthesissafety,synthesisresilient,BARTOCCI2021229,AnandMNS23} is that they did not focus on defining and proving foundational properties as our theorems (1) to (4), on randomized strategies, the number of Joker moves, determinacy, and admissibility, except that the authors of \cite{AnandMNS23} show that their strategies are incomparable with admissible strategies. 

\paragraph{Probabilities.}
A different setting where strategies are optimized is the area of stochastic games, where an opponent can take nondeterministic choices, but probabilities for the opponent's actions are known. For example, Nachmanson et al. \cite{NachmansonVSTG2004} consider such games with probabilities, and additionally also costs for actions. They optimize this for obtaining tests that cover all edges.

We note that our Joker games deal differently with nondeterminism than stochastic games like Markov Decision Processes (MDP).
Assume the following game, where the opponent has two moves from state $q$: move $x$ leads to the winning state, e.g.,  $q \xrightarrow{x} \smiley$, while the other move $y$ is a self-loop  $q \xrightarrow{y} q$. The MDP approach assumes a probability distribution over the moves $x, y$. If this distribution assigns a non-zero probability to move $y$, then Player 1 can always win this game, and needs no Jokers. However, in Joker games, we do not assume non-zero probabilities, i.e. the opponent may choose to not play $y$ at all at his own initiative. Also, in Joker games, we do not assume that we know probabilities for opponent actions.
Thus, the MDP probability-maximization approach \cite{NachmansonVSTG2004} is different from the Joker approach.

\paragraph{Cooperation.}
Different forms of cooperation are considered in related work. For example, 
in controller synthesis: \cite{ChatterjeeH07-assume-guarantee}  the synthesis problem
is refined into a co-synthesis problem, where 
  components 
  compete conditionally: their first aim is to satisfy their own specification, and their second aim is to violate the other component's specification.
In \cite{DBLP:conf/atva/BloemEK15}, a hierarchy of collaboration levels is defined, formalizing how cooperative a controller for an LTL specification can be.
Furthermore, Almagor et al. \cite{AlmagorK20} define several notions of hopefulness, to perform `good-enough' synthesis of strategies.
Finally, Kupferman et al. \cite{KupfermanS2023} consider games where players can pay each other to perform a preferred action.

\section{Concurrent Games}
\label{sec:games}

We consider concurrent games played  by two players on a game graph. 
In each state, Player 1 and 2 \emph{concurrently} choose an action, leading the game in one of the (nondeterministically chosen) next states.

\begin{defi}\label{def:gamearena}
 A \emph{concurrent game} is a tuple $\game=(Q,q^0,\act_1,\act_2,\allowbreak\Gamma_1,\allowbreak\Gamma_2, \allowbreak\moves)$ where:
 \begin{itemize}
  \item $Q$ is a finite set of states,
  \item $q^0 \in Q$ is the initial state,
  \item For $i\in \{1,2\}$, $\act_i$ is a finite and non-empty set of Player $i$ actions,
  \item For $i\in \{1,2\}$, $\Gamma_i: Q\to 2^{\act_i} \setminus \emptyset$
  is an enabling condition, which assigns to each state $q$ a non-empty set $\Gamma_i(q)$ of actions available to Player $i$ in $q$, 
  \item $\moves: Q \times \act_1 \times \act_2 \rightarrow 2^Q$ is a function that
 given the actions of Player 1 and 2 determines the set of next states $Q' \subseteq Q$ the game can be in.
 We require that $\moves(q,a,x) = \emptyset$ iff $a\not\in\Gamma_1(q) \vee x\not\in\Gamma_2(q)$.
  \end{itemize}
\end{defi}

\begin{figure}[b]
\centering
    \begin{tikzpicture}[shorten >=1pt,node distance=2cm,>=stealth']
     	\tikzstyle{every state}=[draw=black,text=black,inner sep=1pt,minimum size=11pt,initial text=]
 	\node[state,fill=lightgray] (1) {2};
 	\node[state,fill=lightgray] (2) [below of=1] {4};
 	\node[state,fill=lightgray] (3) [left of=2] {3};
 	\node[state,initial,initial where=left,fill=lightgray] (4) [left of=1] {1};
 	\node[circle,inner sep=-2pt,fill=lightgray] (5) [right of=1] {\LARGE\smiley};
 	\node[circle,inner sep=-2pt,fill=lightgray] (6) [below left=1cm of 1] {\LARGE\frownie};
 	\path[->]
 	(1) edge node [above] {$a$ $x$} (5)
 	(1) edge node [below right] {$a$ $y$} (6)
 	(1) edge [bend left] node [right] {$b$ $x$} (2)
    (1) edge [loop above] node [above right=-2mm and 0mm] {$b$ $y$} (1)
 	(2) edge [bend right] node [below right] {$a$ $x$} (5)
 	(2) edge node [above right] {$a$ $x$} (6)
 	(3) edge node [above] {$a$ $y$} (2)
 	(3) edge node [above left] {$a$ $x$} (6)
 	(4) edge [bend right] node [left] {$b$ $x$} (3)
 	(4) edge node [above] {$a$ $x$} (1)
    (5) edge [loop right] node [right] {$a$ $x$} (5)
 	(6) edge [loop left] node [left] {$a$ $x$} (6)
 	;
    \end{tikzpicture}
    \caption{Concurrent game $\gameaorb$ with states $\{1,2,3,4,\smiley,\frownie\}$, Player 1 actions $\{a,b\}$, Player 2 actions $\{x,y\}$, and initial state 1. 
   {The $\moves$ function and the enabling conditions $\Gamma_1$, and $\Gamma_2$ of $\gameaorb$ are represented by the edges. For example, the edge from state 1 to 2 indicates that $a \in \Gamma_1(1)$, $x \in \Gamma_2(1)$, and $2 \in \moves(1,a,x)$.}}
    \label{fig:runningexample}
\end{figure}
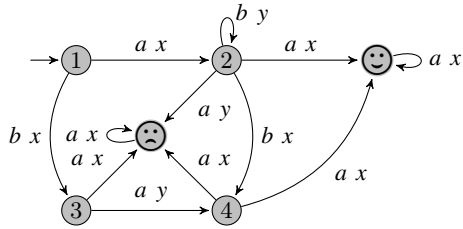

For the rest of the paper, we fix concurrent game $\game = (Q,q^0,\allowbreak\act_1,\act_2,\allowbreak\Gamma_1,\allowbreak\Gamma_2, \moves)$. We will use game $G$ in all definitions and theorems. Furthermore, we will use game $\gameaorb$ from \autoref{fig:runningexample} as running example for this and next section.

Next, we define a \emph{play} as an
infinite sequence of states and actions, and a \emph{play prefix} as a finite prefix of a play. 
An example of a play prefix in the game $\gameaorb$ is: $\pi_{\smiley} = 1 \langle a, x \rangle 2 \langle b, x \rangle 4 \langle a, x \rangle \smiley$.

\begin{defi} \label{def:play}
 An \emph{infinite play} is an infinite sequence\\
 $$\pi = q_0 \langle a_0,x_0 \rangle q_1 \langle a_1, x_1 \rangle  q_2 \dots $$
 with $a_j\in\Gamma_1(q_j)$, 
 $x_j\in\Gamma_2(q_j)$, and 
$q_{j+1} \in \moves({q_j},\allowbreak{a_j},{x_j})$ for all $j \in \mathbb{N}$.
We write $\pi_j^q = q_j$, $\pi_j^a = a_j$, and $\pi_j^x = x_j$ for the $j$-th state, Player 1 action, and Player 2 action respectively.
The set of infinite plays with $\pi_0^q = q$ is denoted $\Pi^\infty(q)$.
We define $\Pi^\infty(\game) = \Pi^\infty(q^0)$.
\end{defi}

\begin{defi}
A  \emph{play} $\pi_{0:j} = q_0\langle a_0,x_0 \rangle q_1 \dots \langle a_{j-1},x_{j-1} \rangle q_j$ is a (finite) play prefix of infinite play $\pi$.
We write $\pi^a_{end}=a_{j-1}$
for the last Player 1 action of the play prefix.

 The set of all (finite) plays of a set of infinite plays $P \subseteq \Pi^\infty(q)$ is denoted $\pref(P) = \{\pi_{0:j} \mid \pi \in P, j \in \mathbb{N}\}$.
 We define $\playpref(\game) = \pref(\Pi^\infty(\game))$, and for any $q \in Q$: $\Pi(q) = \pref(\Pi^\infty(q))$.
 
 \end{defi}
 \noindent
 For the rest of the paper, we fix $R\subseteq Q$ as the set of goal states. Player 1's objective is to reach a state in $R$. In the examples of this paper, we use a state {\smiley} as the (single) goal state, i.e. $R = \{\smiley\}$.
 
 As defined below, a play (prefix) is winning if it reaches some state in $R$; 
 its winning index records
 the first time the play visits $R$. 
 In game $\gameaorb$, the play prefix 
 mentioned before $\pi_{\smiley}= 1 \langle a, x \rangle 2 \langle b, x \rangle 4 \langle a, x \rangle \smiley$ is clearly winning, and has winning index 4.

 \begin{defi} \label{def:winning}
 A play $\pi \in \Pi^\infty(q) \cup \Pi(q)$ for $q \in Q$ is {\em winning} for reachability goal $R \subseteq Q$, if there exist a $j \in \mathbb{N}$ such that $\pi_j^q \in R$.
 A play prefix $\pi_{0:j}$ is winning, if there exists a $k \in \mathbb{N}$ such that $k \le j$ and $\pi_k^q \in R$.
 The \emph{winning index} of a play $\pi \in \Pi^\infty(q) \cup \Pi(q)$ 
 is the first index where $R$ is reached:\\
  $\ind(\pi,\finally) = \min\{j \in \mathbb{N} \mid \pi^q_j \in R\}$, where $\min{\emptyset} = \infty$.
 \end{defi}
 
Players choose their actions according to a \emph{strategy}: given the play until now, each player determines their next move. 
Below, we define deterministic strategies;  randomized ones are deferred until 
 Section~\ref{sec:properties-joker-games}.

A strategy is \emph{positional}, if the choice for an action only depends on the last state of the play.
A positional Player 1 strategy for game $\gameaorb$ is a strategy where Player 1 chooses $b$ in state 1, and $a$ in other states.
The game outcomes for a given Player 1 strategy, are the possible plays in the game against any Player 2 strategy.

\begin{defi} \label{def:strategy}
 A \emph{strategy} for Player $i \in \{1,2\}$ starting in state $q \in Q$ is a function $\sigma_i: \Pi(q) \rightarrow \act_i$, such that $\sigma_i(\pi) \in \Gamma_i(\pi^q_{end})$ for any $\pi \in \Pi(q)$.
 We write $\Sigma_i(q)$ for the set of all Player $i$ strategies starting in $q$, and set $\Sigma_i(\game) = \Sigma_i(q_0)$.
 
  A strategy $\sigma_i \in \Sigma_i(q)$ is \emph{positional} if for all plays 
  $\forall \pi,\tau \in \Pi(q)$ we have $\pi^q_{end} = \tau^q_{end} \implies \sigma_i(\pi) = \sigma_i(\tau).$

The \emph{outcome} of a Player 1 strategy $\sigma_1 \in \Sigma_1(q)$ is the set of infinite plays that occur if Player 1 plays according to $\sigma_1$:
\[\Outc(\sigma_1) =  \{ \pi \in \Pi^\infty(q) \mid \forall j \in \mathbb{N}: \   
\sigma_1(\pi_{0:j}) = \pi_{j+1}^a \}\]
\end{defi}

A Player 1 strategy is \emph{winning} if all its game outcomes are winning.
The \emph{winning region}
collects all states where 
Player 1 can win, i.e., from which she has a winning strategy.
In $\gameaorb$, Player 1 has no winning strategy from any state except $\smiley$.

\begin{defi} \label{def:winstrat}
Let $q \in Q$ be a game state.
A Player 1 strategy $\sigma_1 \in \Sigma(q)$ is \emph{winning}, if all plays from $\Outc(\sigma_1)$ are winning.
The Player 1 \emph{winning region} $\winrop(\game,R)$
for game $\game$ and goal $R$ is the set of all states $Q' \subseteq Q$ such that for each $q \in Q'$, Player $1$ has a winning strategy $\sigma_1 \in \Sigma_1(q)$.
\end{defi}

\section{Joker games}
\label{sec:joker-games}

\subsection{Joker games as cost games}

Joker strategies formalize the notion of best-effort strategy by allowing Player 1 to choose,
apart from her regular moves, a so-called Joker move. 
We associate to each concurrent game $\game$ a Joker game $\jokergame$. Joker strategies are (regular) strategies in $\jokergame$. 
With a Joker move, Player 1 chooses both a move $a$ of her own, and a move $x$ of Player 2, as well as the next state reached when executing these moves. In this way, Player 1 controls both Player 2 and the nondetermististic choice that determines the next state $q' \in \moves(q,a,x)$.
In the Joker game, a Player 1 strategy may use both Player 1 moves and Joker moves.

Since we are interested in strategies where Player 1 needs a minimum number of Jokers to win, we associate cost 1 to Joker moves, and cost 0 to other moves. 
Cost-minimal strategies are then the strategies that are  most independent from the opponent, but defined from non-winning states.
Below, $\jokersymb$ refers to parts of a Joker game that differ from the regular game $G$, and 
 $\jokersymbreal$ to states, moves, etc. where Jokers are actually played. 
\begin{defi} \label{def:costgame}
 We associate to each concurrent game $\game$ a Joker game 
 $\jokergame=(Q,q^0,\act_1 \cup (\act_1 \times \act_2 \times Q),\act_2,\Gamma_1^\jokersymb,\Gamma_2, \moves^\jokersymb,\cost)$ where:
 \begin{align*}
 &\Gamma_1^\jokersymbreal(q) = \{(a,x,q') \in \Gamma_1(q) \times\Gamma_2(q) \times Q \mid q' \in \moves(q,a,x) \}\\
  &\Gamma_1^\jokersymb(q) = \Gamma_1(q) \cup \Gamma_1^\jokersymbreal(q)\\
  &\moves^\jokersymb (q,a,x) = \begin{cases}
                    \{q'\} &\text{if } a=(a',x',q') \in \Gamma_1^\jokersymbreal(q)\\
                    \moves(q,a,x) &\text{otherwise}                    
                   \end{cases}\\                
 &\cost(q,a) = \begin{cases}
               1 &\text{if } a \in \Gamma_1^\jokersymbreal(q)\\
               0 &\text{otherwise}\\
   %            \text{undefined} &\text{otherwise}
              \end{cases}                   
 \end{align*}
 We write $\Sigma_1^\jokersymb(q)$ to denote the set of strategies in $\jokergame$ starting at state $q$.
 \end{defi}

As an example, consider the Joker game of $\gameaorb$. We then have that $(a,x,\smiley) \in \Gamma_1^\jokersymbreal(2)$, and that $\costop(2,(a,x,\smiley)) = 1$.

For the rest of the paper we fix Joker game $\jokergame$. 
In the remainder of the paper we often use the following two observations about \autoref{def:costgame}:
\begin{enumerate}
    \item The states of a concurrent game are the same as its associated Joker game.
    \item  All plays of $\game$ are also plays of $\jokergame$. 
\end{enumerate}

 \autoref{def:costfunction} defines the cost of plays, strategies, and states.
 The cost of a play $\pi$ arises by adding the costs of all moves until the goal $R$ is reached. If $R$ is never reached, the cost of $\pi$ is $\infty$. 
 The cost of a strategy $\sigma_1$ considers the worst case resolution of Player 2 actions and nondeterministic choices.
 The cost of a state $q$ is the cost of Player 1's cost-minimal strategy from $q$.

 For example, in \autoref{fig:runningexample}, we have $\costop(1 \langle a,x \rangle 2 \langle a,x \rangle \smiley) = 0$.
 However, against a Player 1 strategy $\sigma_1$ that chooses Player 1 action $a$ from both state 1 and 2, Player 2 has a strategy that chooses $y$ in state 2. The game then progresses to $\frownie$, from where state $\smiley$ can never be reached, so the Player 1 strategy $\sigma_1$ has cost $\infty$. On the other hand, if Player 1 uses the strategy $\sigma_1^\jokersymb$ that chooses Player 1 action $a$ from state 1, and Joker action $(a,x,\smiley)$ from state 2, the game progresses to state $\smiley$ in two moves, no matter what strategy Player 2 uses. Hence, the cost of strategy $\sigma_1^\jokersymb$ is 1.

 \begin{defi} \label{def:costfunction}
 Let $q \in Q$ be a state, $\pi \in \Pi^\infty(q) \cup \Pi(q)$ a play, and $\sigma_1 \in \Sigma_1^\jokersymb(q)$ a strategy in Joker game $\jokergame$. For goal states $\finally$, define their cost as follows:
 \begin{align*}
 \cost(\pi) & = 
 \begin{cases}
  \Sigma^{{\ind(\pi,\finally)}-1}_{j=0} \cost(q_j,a_j) & \text{if } \pi \text{is winning}\\ %\quad\text{ (note: $\Sigma$ denotes summation)}\\
  \infty & \text{otherwise}\\
 \end{cases}\\
 %&\cost(\pi,\finally) = \cost(\pi_{0:\ind(\pi,\finally)})\\
 \cost(\sigma_1) & = 
 \sup_{\pi \in \Outc(\sigma_1)} \cost(\pi)\\
 \cost(q) & = \inf_{\sigma_1 \in \Sigma_1^\jokersymb(q) } \cost(\sigma_1)
  \end{align*}
 A strategy $\sigma \in \Sigma_1^\jokersymb(q)$ is  \emph{cost-minimal} if 
 $\sigma \in \underset{\sigma_1 \in \Sigma_1^\jokersymb(q)}{\mathrm{arg inf}} \{\cost(\sigma_1) \}$
\end{defi}

\subsection{Winning states in Joker games}
For cost games, there is a standard fixed point computation to compute the minimum cost for reaching a goal state from state $q$. 
This  \nolinebreak com\-putation relies on the following equations, where $v_n$ denotes the cost computed in iteration \nolinebreak $n$:
\begin{align*}
 v_0(q) & = 0  \text{ if $q\in R$} \\
 v_0(q) & = \infty  \text{ if $q\notin R$}\\
 v_{k+1}(q) & = \min_ {a\in\Gamma_1(q)}\max_{x\in\Gamma_2(q)}\max_{q'\in\moves(q,a,x)} \cost(q,a) + v_k (q')
\end{align*}
For finite games, a fixed point will be reached where $v_N = v_{N+1}$, for which it holds that $v_N(q) = \cost(q)$ for all $q$.

\paragraph{Winning Joker games via attractors.}
Joker games allow the fixed point computation to be simplified, by exploiting their specific structure,
with all costs being either  1 for Joker actions, or 0 otherwise.
We adapt the classical attractor construction (see e.g., \cite{zielonka}) on the original game $\game$. Though we compute winning states for the Joker game, this attractor can be computed on the concurrent game.

The construction relies on two concepts: 
the predecessor $\pre(Q')$ contains all states with some move into  $Q'$; 
the controllable predecessor $\cpre_1(Q')$ contains those states where Player 1 can force the game into $Q'$, no matter how Player 2 plays and how the nondeterminism is resolved. 
We note that $\pre(Q')$ can be equivalently defined as the states $q \in Q$ with $(a,x,q') \in \Gamma_1^\jokersymbreal(q)$ (for $q'\in Q'$). Hence, via $\pre(Q')$, we consider all Joker moves into $Q'$.
For example, in game $\gameaorb$, we have  $\pre(\{\smiley\}) = \{2,4\}$, and $\cpre_1(\{\smiley\}) = \emptyset$.

\begin{defi}
\label{def:predecessors}
 Let $Q' \subseteq Q$ be a set of states.
The \emph{predecessor} $\pre(Q')$ of $Q'$, and the Player 1 \emph{controllable predecessor} $\cpre_1(Q')$ of $Q'$ are:
\begin{align*}
  \pre(Q')  & = \{q \in Q \mid \exists a \in \Gamma_1(q), \exists x \in \Gamma_2(q), \exists q' \in \moves(q,a,x): q' \in Q'\}\\
   \cpre_1(Q') &  = \{q \in Q \mid \exists a \in \Gamma_1(q), \forall x \in \Gamma_2(q): \moves(q,a,x) \subseteq Q'\}
\end{align*}
\end{defi}

The classical \emph{attractor} is the set of states from which Player 1 can force the game to reach $R$, winning the game. 
It is constructed by expanding the goal states via $\cpre_1$, until a fixed point is reached \cite{dAHK98};
the rank $k$ indicates in which computation step a state was added \cite{zielonka}. 
Thus, the lower $k$, the fewer moves Player 1 needs to reach its goal.
As example, observe that $\attrop(\gameaorb,\{\smiley\}) = \{\smiley\}$.

\begin{defi} \label{def:attr}
 The Player 1 \emph{attractor} is $\attrf$, where:
 \begin{align*}
  &\attrfi{0} = R\\
  &\attrfi{k+1} = \attrfi{k}\ \cup \cpre_1(\attrfi{k})\\
  &\attrf = \bigcup_{k \in \mathbb{N}} \attrfi{k}
 \end{align*}
  
The function $\arankg: Q \rightarrow \mathbb{N}$ associates to each state $q\in Q$ a rank $\arankg(q) = \min \{k \in \mathbb{N} \mid q \in \attrfi{k}\}$.
Recall that $\min \emptyset = \infty$.
\end{defi}

\begin{figure}[t!]
    \centering
    \begin{tikzpicture}
       \node[ellipse,draw=black,fill=gray!10] (Jn) [minimum width=10.4cm,minimum height=2.8cm] {};
       \node (Jokern) [right=-1.9cm of Jn] {\footnotesize$\jokerg$};
       \node[ellipse,draw=black,fill=gray!20] (D) [left=-8.3cm of Jn,minimum width=8.2cm,minimum height=2.5cm] {};
       \node (dots) [right=-0.7cm of D] {\footnotesize$\dots$};
        \node[ellipse,draw=black,fill=gray!35] (J) [left=-7.3cm of D, minimum width=7.2cm,minimum height=2.2cm] {};
        \node (Joker) [right=-3.5cm of J,text width=2.2cm] {\footnotesize$\jokerreal{1}$ =  $\joker{0}$ $\cup\  \pre(\joker{0})$};
         \node[ellipse,draw=black,fill=gray!50] (A) [left=-3.7cm of J,
         minimum width=3.4cm,minimum height=1.7cm] {};
         \node (Attr) [right=-2.2cm of A,text width=2cm] {\footnotesize$\attrf$ = $\joker{0}$};
         \node[circle,draw=black,fill=gray!80] (R) [left=-1.2cm of A, minimum height=1cm,align=center] {\footnotesize R};
    \end{tikzpicture}
    \caption{Illustration of the Joker attractor computation: it starts with states in R initially and then adds states using the attractor and Joker attractor operations, until a fixpoint is reached.}
    \label{fig:ui}
    \end{figure}
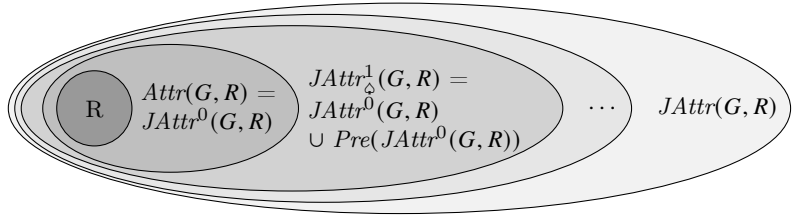

We define the \emph{Joker attractor} to interleave the standard attractor and the predecessor operator. 
\autoref{fig:ui}  illustrates the computation.
Since $\attrop$ and $\pre$ only use elements from $\game$, the computation is performed in $\game$. We will see that for defining the Joker attractor strategy (\autoref{def:jokerstrat}) we do need Joker game $\jokergame$.\
With $\jrankg(q)$ we denote after how many $\pre$ operations $q$ was first added; we will see that this is the number of used Jokers.

\begin{defi} \label{def:jokers}
 The Player 1 Joker attractor is $\jokerg$, where:
 \begin{align*}
  \joker{0} &= \attrf\\
  \jokerreal{k+1} &= \joker{k} \cup \pre(\joker{k})\\
  \joker{k+1} &= \attrop(\game,\jokerreal{k+1})\\
  \jokerg &= \bigcup_{k \in \mathbb{N}} \joker{k}
  \end{align*}
   
 We call $\jokerg$ the Joker attractor of $\game$. The Joker states are 
  $\jstate = \bigcup_{k \in \mathbb{N}} \jokerreal{k+1}\setminus\joker{k}$.
 To each Joker attractor $\jokerg$ we associate a Joker rank function $\jrankg : Q \rightarrow \mathbb{N}$, where for each state $q \in Q$ we define $\jrankg(q) = \min \{k \in \mathbb{N} \mid q \in \joker{k}\}$.
\end{defi}

\begin{exa}
As an example, we show the computation of the Joker attractor for game $\gameaorb$ with goal $\smiley$ below. We also indicate for each state what its $\jrankop$ is, and whether it is a Joker state or not.

\noindent
 \begin{minipage}{0.6\textwidth}
 \begin{align*}
  \jokerop^0(\gameaorb,\smiley) &= \{\smiley\}\\
  \jokerop^{1}_\jokersymb(\gameaorb,\smiley) &= \{2,4,\smiley\}\\
  \jokerop^{1}(\gameaorb,\smiley) &= \{1,2,4,\smiley\}\\
  \jokerop^{2}_\jokersymb(\gameaorb,\smiley) &= \{1,2,3,4,\smiley\}\\
  \jokerop^{2}(\gameaorb,\smiley) &= \{1,2,3,4,\smiley\}\\
  \jokerop^{3}(\gameaorb,\smiley) &= \{1,2,3,4,\smiley\}
 \end{align*} 
  \end{minipage}
 \quad
 \begin{minipage}{0.3\textwidth}
 \centering
\begin{tabular}{c|c|c}
 $q \in Q$ & $\jrankop
 $ & Joker state\\\hline
 1 & 1 & no\\
 2 & 1 & yes\\
 3 & 2 & yes\\
 4 & 1 & yes \\
 \smiley & 0 & no \\
 \frownie & $\infty$ & no\\
\end{tabular}\\
\noindent
 \end{minipage}
 \bigskip
\end{exa}

\autoref{thm:costgameisjokergameNew} states that the cost of state $q$ in $\jokergame$, i.e. the minimum number of Jokers needed from $q$ to reach a goal state, is equal to $\jrankg(q)$.
The proof follows from the following observations:
\begin{itemize}
    \item Clearly, states that can be won by Player 1 in $\game$ have cost 0, so these fall into $\joker{0}$.
    \item Similarly, if all states in $Q'$ can be won with at most $k$ jokers, then so can states in $\attrop(Q')$. 
    \item The \emph{Joker states} of set $\joker{k+1}$, are the states where Player 1 can only win from any opponent if she uses a Joker. By playing a Joker, the game moves to a state in $\joker{k}$. Joker states are the  predecessors of $\joker{k}$.
\end{itemize}
Together, these  observations establish that in all states of $\joker{k}$, the game can be won with at most $k$ Jokers, i.e., with cost $k$.

\begin{restatable}{thm}{jokeriscost}
\label{thm:costgameisjokergameNew}
For all $q \in Q$, we have
$\jrankg(q)=\cost(q)$.
\end{restatable}
\prooflink{proofthm1}

\subsection{Winning strategies in Joker games}
 The attractor construction cannot only be used to compute the states of the Joker attractor, but also to construct the corresponding winning strategy. 
 As is common in strategy construction, we do so by adding some extra administration to the (controllable) predecessor operations $\pre$ and $\cpre$. 
The witnessed predecessor $w\pre(Q')$ returns the states \emph{and} their Joker moves into $Q'$.
Similarly, the witnessed controllable predecessor $w\cpre(Q')$ returns the states, \emph{and} their Player 1 actions that force the game into $Q'$.

We have seen that in game $\gameaorb$, $\pre(\{\smiley\}) = \{2,4\}$. With $w\pre$ we compute the corresponding Joker moves of these states: $w\pre(\{\smiley\}) = \{(2,(a,x,\smiley)),(4,(a,x,\smiley))\}$. And instead of $\cpre_1(\{2\}) = \{1\}$ we compute the controlled predecessor with Player 1 actions: $w\cpre_1(\{2\}) = \{(1,a)\}$.

 \begin{defi}
      Let $Q' \subseteq Q$ be a set of states.
The \emph{witnessed predecessor} $w\pre(Q')$ of $Q'$, and the Player 1 \emph{witnessed controllable predecessor} $w\cpre_1(Q')$ of $Q'$ are:
 \begin{align*}
    w\pre(Q')  & = \{(q,(a,x,q')) \in Q \times (\act_1 \times \act_2 \times Q)  \mid  q' \in \moves(q,a,x) \cap Q'\}\\
    w\cpre_1(Q') &  = \{(q,a) \in Q \times \act_1 \mid a \in \Gamma_1(q) \wedge (\forall x \in \Gamma_2(q): \moves(q,a,x) \subseteq Q')\}
\end{align*}
 \end{defi}

 Now, we obtain the sets of winning actions, by replacing, in the construction of  $\jokerg$, $\pre$ and $\cpre$ by 
 $w\pre$ and $w\cpre_1$, respectively. 
 We only select actions for moving from newly added states of the $k+1$-th attractor, or the $k+1$-th Joker attractor set, respectivly, to states of the $m$-th attractor, or $m$-th Joker attractor set, respectively, such that $m \neq k$. This way, always a state of a previous attractor or previous Joker set is reached, ensuring progress towards reaching a goal state.

 \begin{exa}
 For game $\gameaorb$ we compute the following witnessed Joker attractor sets:
  \begin{align*}
  w\jokerop^0(\gameaorb,\{\smiley\}) &= \emptyset\\
  w\jokerop^1_\jokersymb(\gameaorb,\{\smiley\}) &= \{(2,(a,x,\smiley)),(4,(a,x,\smiley))\}\\
  w\jokerop^{1}(\gameaorb,\{\smiley\}) &= \{(1,a),(2,(a,x,\smiley)),(4,(a,x,\smiley))\}\\
  w\jokerop^{2}_\jokersymb(\gameaorb,\{\smiley\}) &= \{(1,a),(2,(a,x,\smiley)),(3,(a,x,4)),(4,(a,x,\smiley))\}\\
 %&=   w\jokerop^{2}(\gameaorb,\{\smiley\}) = \jokerop^{3}(\gameaorb,\{\smiley\})
 \end{align*} 
 \end{exa}

\pagebreak[5]

 \begin{defi} \label{def:witness}
We define the \emph{witnessed} attractor $w\attrf$ and \emph{witnessed} Joker attractor $w\jokerg$:

  \begin{align*}
  w\attrfi{0} &= \emptyset\\
  w\attrfi{k+1} &= w\attrfi{k} \cup \{(q,a) \in w\cpre_1(\attrfi{k}) \mid q \notin \attrfi{k}\}\\
  w\attrf &= \bigcup_{k \in \mathbb{N}}w\attrfi{k}\\
  w\joker{0}&= \emptyset\\
  w\jokerreal{k+1}&= w\joker{k} \cup \{ (q,(a,x,q')) \in w\pre(\joker{k}) \mid q \notin \joker{k} \}\\
  w\joker{k+1} &=  w\attrop(\game,w\jokerreal{k+1})\\
  w\jokerg &= \bigcup_{k \in \mathbb{N}}w\joker{k}
  \end{align*}
  \end{defi}

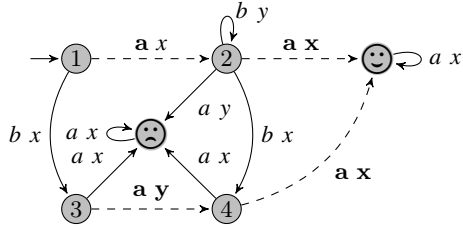
\begin{figure}[ht!]
    \centering
    \begin{tikzpicture}[shorten >=1pt,node distance=2cm,>=stealth']
     	\tikzstyle{every state}=[draw=black,text=black,inner sep=1pt,minimum size=11pt,initial text=]
 	\node[state,fill=lightgray] (1) {2};
 	\node[state,fill=lightgray] (2) [below of=1] {4};
 	\node[state,fill=lightgray] (3) [left of=2] {3};
 	\node[state,initial,initial where=left,fill=lightgray] (4) [left of=1] {1};
 	\node[circle,inner sep=-2pt,fill=lightgray] (5) [right of=1] {\LARGE\smiley};
 	\node[circle,inner sep=-2pt,fill=lightgray] (6) [below left=1cm of 1] {\LARGE\frownie};
 	\path[->]
 	(1) edge [dashed] node [above] {\textbf{a} \textbf{x}} (5)
 	(1) edge node [below right] {$a$ $y$} (6)
 	(1) edge [bend left] node [right] {$b$ $x$} (2)
    (1) edge [loop above] node [above right=-2mm and 0mm] {$b$ $y$} (1)
 	(2) edge [bend right,dashed] node [below right] {\textbf{a} \textbf{x}} (5)
 	(2) edge node [above right] {$a$ $x$} (6)
 	(3) edge [dashed] node [above] {\textbf{a} \textbf{y}} (2)
 	(3) edge node [above left] {$a$ $x$} (6)
 	(4) edge [bend right] node [left] {$b$ $x$} (3)
 	(4) edge [dashed] node [above] {\textbf{a} $x$} (1)
 	(5) edge [loop right] node [right] {$a$ $x$} (5)
    (6) edge [loop left] node [left] {$a$ $x$} (6)
 	;
    \end{tikzpicture}
    \caption{Game $\gameaorb$, as in \autoref{fig:runningexample}, but with dashed edges for moves of the Joker attractor strategy. On these dashed edges, if both the Player 1 and the Player 2 action are bold, then a Joker action is played. Otherwise, if only the Player 1 action is bold, then a (normal) Player 1 action is played. Note that non-dashed moves will never be played with this Joker strategy, no matter what Player 2 strategy or resolution of nondeterminism is used. Also, the dashed move from state 3 will not be played, because the strategy does not go there from the initial state.
    }
    \label{fig:jokergame}
\end{figure}

 A {\em Joker attractor strategy} in  $\jokergame$ is a strategy that 
 plays according to the witnesses: in Joker states, a Joker action from $w\jokerg$ is played. In non-Joker states $q$, the strategy takes its action from an attractor witness $w\attrf$ that is also included in $w\jokerg$. The Joker attractor strategy is not defined in states not included in $\jokerg$, since no goal state can be reached anymore (as we will see in \autoref{thm:jokerwinreach}). It is also undefined in goal states $\finally$, where the goal has already been reached.
 \autoref{fig:jokergame} shows the Joker strategy of game $\gameaorb$.

  \begin{defi}\label{def:jokerstrat}
  A strategy $\sigma_1 \in \Sigma_1^\jokersymb(q)$ in $\jokergame$ is a \emph{Joker attractor strategy}, if for any $\pi \in \playpref(\jokergame)$ 
  we have  $\pi^q_{end} \in \jokerg\setminus\finally \implies (\pi^q_{end},\sigma_1(\pi)) \in  w\jokerg$.
\end{defi}

\section{Properties of Joker games}
\label{sec:properties-joker-games}

We establish four fundamental properties of Joker games. 

\begin{figure}[b!]
\centering
\begin{tikzpicture}[shorten >=1pt,node distance=1.3cm,>=stealth']
\tikzstyle{every state}=[draw=black,text=black,inner sep=1pt,minimum size=11pt,initial text=]
 	\node[state,fill=lightgray,initial,initial where=left] (1) {1};
 	\node[state,fill=lightgray] (2) [below right of=1] {2};
 	\node[circle,inner sep=-2pt,fill=lightgray] (3) [below left of=1] {\LARGE\smiley};
 	\node[circle,inner sep=-2pt,fill=lightgray] (4) [right of=2] {\LARGE\frownie};
 	\path[->]
 	(1) edge [dashed] node [above right] {$a$ $y$} (2)
 	(1) edge [dotted] node [above left] {$a$ $x$} (3)
 	(2) edge [dashed] node [above] {$a$ $x$} (3)
 	(2) edge [dashed] node [above] {$a$ $y$} (4)
 	(3) edge [loop left] node [left] {$a$ $x$} (3)
 	(4) edge [loop right] node [right] {$a$ $y$} (4)
 	;
 \end{tikzpicture}
 \caption{A cost-minimal strategy is depicted by the dashed edges. It plays a cost-0 $a$ action in state 1, and hopes Player 2 plays $x$ to arrive in {\smiley} with cost 0. If Player 2 plays $y$, the strategy plays a Joker in state 2 to win nevertheless. The computation of the Joker attractor yields that state 1 and 2 both have $\jrankop$ 1. Since initial state 1 is always reached first by the predecessor, the Joker attractor strategy (dotted) plays a Joker in state 1 immediately, and reaches {\smiley} directly from state 1. This example shows that cost-minimal strategies from state 1 may use fewer than $\jrankg(1)$ Jokers (against some but not all opponents), and that Joker attractor strategies from state 1 always use $\jrankg(1)$ Jokers.}
 \label{fig:costless}
\end{figure}
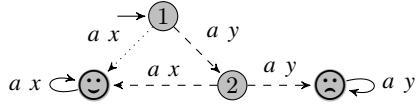

\paragraph{1. All outcomes use same the number of Jokers.}
A first, and perhaps surprising result (\autoref{thm:nrjokermoves}) is that all outcomes of a Joker attractor strategy use exactly the same number of Joker actions: all plays starting in state $q$ contain  exactly $\jrankg(q)$ Jokers. This property follows by induction from the union-like computation illustrated in \autoref{fig:ui}.
Cost-minimal strategies in general cost games do not have this property, because some outcomes may use lower costs. In particular, cost-minimal strategies that are not obtained via the attactor construction may also lead to plays which differ in the number of Jokers they use. This is illustrated in  \autoref{fig:costless} and shows that attractor strategies are special among the cost-minimal strategies.

\bigskip\bigskip
\begin{restatable}{thm}{nrjokermoves}
\label{thm:nrjokermoves}
Let $q \in \jokerg$.
Then: 
\begin{enumerate}
    \item Let $\sigma^J_1 \in \Sigma_1^\jokersymb(q)$ be a Joker attractor strategy in $\jokergame$.
    Then any play $\pi \in \Outc(\sigma^J_1)$ has \emph{exactly} $\jrankg(q)$ Joker actions in winning prefix $\pi_{0:\ind(\pi)}$.
    \item Let $\sigma_1 \in \Sigma_1^\jokersymb(q)$ be a cost-minimal strategy in $\jokergame$.
    Then any play $\pi \in \Outc(\sigma_1)$ has \emph{at most} $\jrankg(q)$ Joker actions in the winning prefix $\pi_{0:\ind(\pi)}$.
\end{enumerate}
\end{restatable}
\prooflink{proofthm2}\bigskip

The intuition of the proof of \autoref{thm:nrjokermoves} is illustrated in \autoref{fig:costless}: a Joker is always played in a Joker state. By construction of the Joker attractor strategy, the Joker action moves the game from a state $q$ to a state $q'$, such that the $\jrankop$ of $q$ is one higher than the $\jrankop$ of $q'$.
In non-Joker states no Joker is used to reach a next Joker state.

\paragraph{2. Characterization of winning states of Joker games.} We establish that:
(1) for every goal $R$, the winning region in Joker game $\jokergame$ (i.e., states where Player 1 can win with any strategy, not necessarily cost-minimal) coincides with the Joker attractor $\jokerg$, and
(2) the set of all states having a play reaching a state from $R$ coincide with Joker attractor $\jokerg$.
The intuition for this is that, by playing a Joker in any state, Player 1 can force the game to any state that is reachable. Consequently, states from where a goal state can be reached, are winning, and in the Joker attractor.

The states from where Player 1 cannot ever win the game, because no goal state is reachable, is thus the set $\jokerg \setminus R$. In the examples of this paper we represent these states with the single state {\frownie}.

 \begin{restatable}{thm}{jokerwinreach}
 \label{thm:jokerwinreach}
 Let $\reach = 
 \{q \in Q \mid q \text{ can reach a state } q'\in R\}$.
 Then
 \begin{align*}
     \winrop(\jokergame,R) = \jokerg = \reach
 \end{align*}
 \end{restatable}
 \prooflink{proofthm3}

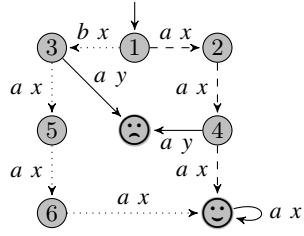
\begin{figure}[b!]%[9]
  \centering
     \begin{tikzpicture}[shorten >=1pt,node distance=1.1cm,>=stealth']
     	\tikzstyle{every state}=[draw=black,text=black,inner sep=1pt,minimum size=11pt,initial text=]
 	\node[state,fill=lightgray,initial,initial where=above] (1) {1};
 	\node[state,fill=lightgray] (2) [right of=1] {2};
 	\node[state,fill=lightgray] (3) [left of=1] {3};
 	\node[state,fill=lightgray] (4) [below of=2] {4};
 	\node[state,fill=lightgray] (5) [below of=3] {5};
 	\node[state,fill=lightgray] (6) [below of=5] {6};
 	\node[circle,inner sep=-2pt,fill=lightgray] (7) [below of=4] {\LARGE\smiley};
 	\node[circle,inner sep=-2pt,fill=lightgray] (8) [below of=1] {\LARGE\frownie};
 	\path[->]
 	(1) edge [dashed] node [above] {$a$ $x$} (2)
 	(1) edge [dotted] node [above] {$b$ $x$} (3)
 	(2) edge [dashed] node [left] {$a$ $x$} (4)
 	(3) edge [dotted] node [left] {$a$ $x$} (5)
 	(3) edge node [above right=-1mm and -1mm] {$a$ $y$} (8)
 	(4) edge node [below] {$a$ $y$} (8)
 	(4) edge [dashed] node [left] {$a$ $x$} (7)
 	(5) edge [dotted] node [left] {$a$ $x$} (6)
 	(6) edge [dotted] node [above] {$a$ $x$} (7)
    (7) edge [loop right] node [right] {$a$ $x$} (7)
 	;
    \end{tikzpicture}
     \caption{This game $\gamedist$ has a cost-minimal strategy (dashed; cost 1 for the Joker used in state 4) that is not a Joker attractor strategy.  
     For this game, there is a unique Joker attractor strategy (dotted). It selects Player 1 action $b$ from state 1 (to Joker state 3), since states 1 and 2 are added at the same iteration of the attractor over the set with Joker states 3 and 4. The witnessed attractor in iteration $k$ only selects Player 1 actions to states from an iteration $\neq k$.
     The dashed cost-minimal strategy requires fewer moves to reach {\smiley} than the dotted Joker strategy, namely 3 versus 4 moves.
     }
     \label{fig:costminnojokerstrat}
 \end{figure}

\paragraph{3. Joker attractor strategies and minimality.}
\autoref{thm:jokerstratwinning} states correctness of Joker attractor strategies: they are indeed cost-minimal.
The converse, cost-minimal strategies are Joker attractor strategies, is not true, as shown by \autoref{fig:costless} and \ref{fig:costminnojokerstrat}. The game of
\autoref{fig:costminnojokerstrat} also shows that Joker attractor strategies need not take the shortest route to the goal (see also \autoref{sec:short}). Note that a consequence of \autoref{thm:jokerstratwinning} is that Joker strategies are indeed winning in Joker games.

 \begin{restatable}{thm}{jokerstratcostminimal}
 \label{thm:jokerstratwinning}
     Any Joker attractor strategy is cost-minimal (a), but not every cost-minimal strategy is a Joker attractor strategy (b).
 \end{restatable}

\prooflink{proofthm4}
 
\paragraph{4. Determinacy.}
Unlike concurrent games, Joker games are \emph{determined}. That is, in each state of the Joker game, either Player 1 has a winning strategy or Player 2 can keep the game outside $R$ forever. Determinacy (\autoref{thm:jokerdetermined}) follows from \autoref{thm:jokerwinreach}: by $\winrop(\jokergame,\finally) = \jokerg$ we have a winning strategy in any state from $\jokerg$, and by $\jokerg = \reach$ we have that states not in $\jokerg$ have no play to reach $\finally$, so Player 2 wins with any strategy.

\begin{restatable}{thm}{jokerdetermined}
\label{thm:jokerdetermined}
Joker games are determined.
\end{restatable}
\prooflink{proofthm5}

\section{Admissible strategies} \label{sec:admissible}

Various papers  \cite{admissibilityingraphs,nonzerosumgames,Faella09} advocate that a player should play \emph{admissible} strategies if there is no winning strategy. 
Admissible strategies (\autoref{def:dominance}) are the maximal elements in the lattice of all strategies equipped with the dominance order $<_d$. Here, a Player 1 strategy $\sigma_1$ dominates strategy $\sigma'_1$, denoted $\sigma'_1 <_d \sigma_1$,
iff whenever $\sigma'_1$ wins from opponent strategy $\rho$, so does $\sigma_1$.
We hence then define dominant and admissible strategies, in concurrent games, using the set of opponent strategies a Player 1 strategy can win from (\autoref{def:dominance}).
Since the  nondeterministic choice for the next state of the game affects whether Player 1 wins, we introduce Player 3 (\autoref{def:playerthree}) for making this choice. Specifically, given a play prefix $\pi$, and moves $a$ and $x$ chosen by Player 1 and 2, respectively, a Player 3 strategy
chooses one of the next states in $\moves(\pi^q_{end},a,x)$.
The set of winning opponent strategies is then defined as the pairs of Player 2 \emph{and} Player 3 strategies. 

\begin{defi} \label{def:playerthree}
%   Let $q \in Q$. 
  A  Player 3 strategy is a function $\sigma_3: \Pi(q) \times \act_1 \times \act_2 \rightarrow Q$, such that $\sigma_3(\pi,a,x) \in \moves(\pi^q_{end},a,x)$ for all $\pi \in \Pi(q), a \in \Gamma_1(\pi^q_{end}), x \in \Gamma_2(\pi^q_{end})$.
 We write $\Sigma_3(q)$ for the set of all Player $3$ strategies from $q$.
 
  The \emph{outcome} $\Outc(\sigma_1,\sigma_2,\sigma_3)$ of strategies $\sigma_1 \in \Sigma_1(q)$, $\sigma_2 \in \Sigma_2(q)$, and $\sigma_3 \in \Sigma_3(q)$ is the single play $\pi \in \Pi^\infty(q)$ such that:
 \[\forall j \in \mathbb{N}: \   
\sigma_1(\pi_{0:j}) = \pi_{j}^a \wedge \sigma_2(\pi_{0:j}) = \pi_{j}^x \wedge \sigma_3(\pi_{0:j},\sigma_1(\pi_{0:j}),\sigma_2(\pi_{0:j})) = \pi_{j+1}^q\]
\end{defi}

\begin{defi} \label{def:dominance}
 The Player 2 and 3 strategy pairs that are winning in $\game$ for a strategy $\sigma_1 \in \Sigma_1(\game)$ are defined as:
 \begin{align*}
 \wins_{2,3}(\sigma_1,\finally) = \{(\sigma_2,\sigma_3) \in \Sigma_2(\game) \times \Sigma_3(\game) \mid \Outc(\sigma_1,\sigma_2,\sigma_3) \in \winp(\game,\finally) \}
 \end{align*}
 
 For any $q \in Q$, a strategy $\sigma_1 \in \Sigma_1(q)$ is \emph{dominated by} a strategy $\sigma_1' \in \Sigma_1(q)$, denoted $\sigma_1 <_d \sigma_1'$, if $\wins_{2,3}(\sigma_1,\finally) \subset \wins_{2,3}(\sigma_1',\finally)$.
 Strategy $\sigma_1 \in \Sigma_1(q)$ is \emph{admissible} if there is no strategy $\sigma_1' \in \Sigma_1(q)$ with $\sigma_1 <_d \sigma_1'$.
\end{defi}

To compare cost-minimal strategies, including Joker attractor strategies, with admissible strategies, we note that cost-minimal strategies are played in Joker games, where Joker actions have full control over the opponent. Admissible strategies, however, are played in regular concurrent games, without Joker actions. 
To make the comparison, \autoref{def:insp} therefore
% \autoref{def:insp} facilitates the translation of Joker attractor strategies to concurrent games, by
associates to any Player 1 strategy $\sigma$ in $\jokergame$, a Joker-inspired strategy $\sigma_{insp}$ in $G$: if $\sigma$ chooses Joker action $(a,x,q)$, then $\sigma_{insp}$ plays Player 1 action $a$.

\begin{defi} \label{def:insp}
Let $\sigma \in \Sigma_1(q)$ be a Player 1 strategy in Joker game $\jokergame$. Define the \emph{Joker-inspired strategy} $\sigma_{insp} \in \Sigma_1(q)$ of $\sigma$  in concurrent game $\game$ for any $\pi \in \Pi(\game)$ as:
\begin{align*}
    \sigma_{insp}(\pi) = \begin{cases}
     a &\text{ if } \sigma(\pi) = (a,x,q)\\
     \sigma(\pi) &\text{ otherwise}
    \end{cases}
\end{align*}
\end{defi}

In the next two subsections, we study whether admissible strategies are always Joker-inspired cost-minimal stratgies, and whether Joker-inspired cost-minimal strategies are always admissible.

\subsection{Admissible strategies not necessarily cost-minimal}

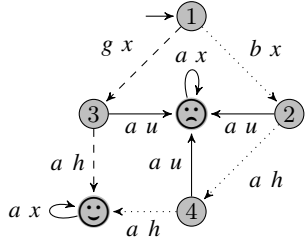
\begin{figure}[ht!]
\centering
     \begin{tikzpicture}[shorten >=1pt,node distance=1.3cm,>=stealth']
	\tikzstyle{every state}=[draw=black,text=black,inner sep=1pt,minimum
	size=11pt,initial text=]
	\node[state,initial,initial where=left,fill=lightgray] (1) {1};
	\node[circle,inner sep=-2pt,fill=lightgray] (5) [below of=1] {\LARGE\frownie};
    \node[state,fill=lightgray] (2) [left of=5] {3};
    \node[circle,inner sep=-2pt,fill=lightgray] (6) [below of=2] {\LARGE\smiley};
	\node[state,fill=lightgray] (3) [right of=5] {2};
    \node[state,fill=lightgray] (4) [below of=5] {4};
	\path[->]
 	(1) edge [dashed] node [above left] {$g$ $x$} (2)
 	(1) edge [dotted] node [above right] {$b$ $x$} (3)
 	(2) edge node [below] {$a$ $u$} (5)
    (2) edge [dashed] node [left] {$a$ $h$} (6)
 	(3) edge [dotted] node [below right] {$a$ $h$} (4)
 	(3) edge node [below] {$a$ $u$} (5)
    (4) edge node [left] {$a$ $u$} (5)
 	(4) edge [dotted] node [below] {$a$ $h$} (6)
    (6) edge [loop left] node [left] {$a$ $x$} (6)
    (5) edge [loop above] node [above] {$a$ $x$} (5)
	;
	\end{tikzpicture}
 \caption{The dotted strategy chooses $b$ in state 1, and after that plays $a$ in the two Joker states 2 and 4. In the Joker game, strategies that pass those two states need to play a Joker action to be winning, since Player 2 can prevent Player 1 from ever reaching {\smiley} by playing a $u$ in state 2 or 4. Hence, the dotted strategy spends two Jokers in the Joker game. The dotted strategy is admissible in the concurrent game though, since it wins from the Player 2 strategy that chooses $u$ in state 3 and $h$ or $x$ anywhere else, while the dashed Joker-inspired strategy looses from this Player 2 strategy.}
 \label{fig:admnonjoker}
\end{figure}

The game of \autoref{fig:admnonjoker} shows that admissible strategies need not be a Joker-inspired strategy of a cost-minimal strategy, as stated by \autoref{thm:admnonjoker}. While cost-minimal strategies ensure that the minimum number of Jokers is used in a Joker game, admissible strategies are not defined to satisfy this condition. The dotted strategy of \autoref{fig:admnonjoker} uses that it already dominates another strategy if there is one opponent strategy in a non-visited state the dotted strategy trivially wins from, while the other looses. Hence, domination does not prevent using too many Jokers in the Joker game.

\begin{restatable}{thm}{admnonjoker}
\label{thm:admnonjoker}
    An admissble strategy is not always a Joker-inspired strategy of a cost-minimal strategy.
\end{restatable}

\prooflink{proofthm6}

\subsection{Admissibility of Joker-inspired, cost-minimal strategies}

\begin{figure}[b!]
\centering
  \begin{tikzpicture}[shorten >=1pt,node distance=1.3cm,>=stealth']
	\tikzstyle{every state}=[draw=black,text=black,inner sep=1pt,minimum
	size=11pt,initial text=]
	\node[state,initial,initial where=above,fill=lightgray] (1) {1};
	\node[state,fill=lightgray] (2) [below of=1] {2};
	\node[circle,inner sep=-2pt,fill=lightgray] (3) [right of=1] {\LARGE\smiley};
	\node[circle,inner sep=-2pt,fill=lightgray] (4) [left of=1] {\LARGE\frownie};
	\path[->]
	(1) edge [dashed] node [above] {$a$ $h$} (3)
	(1) edge node [above] {$a$ $u$} (4)
	(1) edge node [above left] {$a$ $n$} (2)
	(2) edge [dashed] node [below left] {$b$ $x$} (4)
	(2) edge node [below right] {$g$ $x$} (3)
	(3) edge [loop above] node [above] {$a$ $x$} (3)
 	(4) edge [loop left] node [left] {$a$ $x$} (4)
	;
\end{tikzpicture}
  \caption{The dashed strategy uses bad action $b$ in state 2 instead of good action $g$, because using Joker ($a$,$h$,\smiley) in state 1 will make Player 1 win the game. Hence, the Joker-inspired strategy of this winning cost-minimal strategy is not admissible, as it is dominated by strategies choosing $g$ in state~2.}
   \label{fig:nonglobal}
\end{figure}
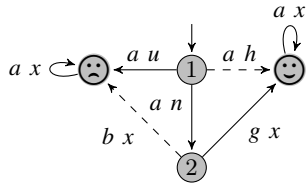

It turns out that Joker-inspired strategies of cost-minimal strategies need not be admissible, for a rather trivial reason: a cost-minimal strategy that chooses a losing move in a non-visited state is not admissible. See \autoref{fig:nonglobal}.
Because actions in non-visited states do not influence the outcome of a game, we want to exclude this case. Therefore, we will only consider the admissibility of \emph{global} cost-minimal strategies (\autoref{def:globcostmin}): strategies that are cost-minimal for any initial state of the Joker game.  We note here that Joker attractor strategies are global-cost-minimal by construction.

\begin{defi} \label{def:globcostmin}
Let $\sigma \in \Sigma_1^\jokersymb(q)$ be a strategy in Joker game $\jokergame$. Then $\sigma$ is \emph{global cost-minimal} if $\sigma$ is  cost-minimal from any $q' \in \reach$.
\end{defi}

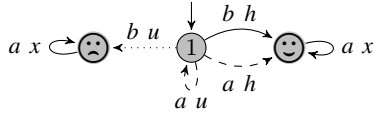
\begin{figure}[t!]
\centering
 \begin{tikzpicture}[shorten >=1pt,node distance=1.3cm,>=stealth']
	\tikzstyle{every state}=[draw=black,text=black,inner sep=1pt,minimum
	size=11pt,initial text=]
	\node[state,initial,initial where=above,fill=lightgray] (1) {1};
	\node[circle,inner sep=-2pt,fill=lightgray] (2) [right of=1] {\LARGE\smiley};
	\node[circle,inner sep=-2pt,fill=lightgray] (3) [left of=1] {\LARGE\frownie};
	\path[->]
  	(1) edge [bend left] node [above] {$b$ $h$} (2)
  	(1) edge [dashed,bend right] node [below] {$a$ $h$} (2)
  	(1) edge [dotted] node [above] {$b$ $u$} (3)
  	(1) edge [loop below, dashed] node [below] {$a$ $u$} (1)
    (2) edge [loop right] node [right] {$a$ $x$} (2)
    (3) edge [loop left] node [left] {$a$ $x$} (3)
	;
	\end{tikzpicture}
 \caption{The dashed Player 1 strategy, that choosing $a$ in state 1, dominates the dotted Player 1 strategy, that choosing $b$ in state 1, because only the first strategy wins from a Player 2 strategy that plays $u$ on the first visit of state 1, but $h$ for all later visits. Both Player 1 strategies are Joker-inspired.}
     \label{fig:memstrat}
\end{figure}

Even when only considering \emph{global} cost-minimal strategies, it turns out some of these strategies may not be admissible when playing against opponent strategies using memory (\autoref{thm:jokeradm}(1)).
\autoref{fig:memstrat} shows an example where a Player 2 strategy uses memory. It has two Player 1 strategies that both need to pass by a Joker state. In this state Player 2 can force the game to the {\frownie} state against one of the Player 1 strategies, such that Player 1 can never win any more. Against the other Player 1 strategy, Player 2 only prevent progress towards {\smiley}, for one move only, via a loop to the same state.

Without memory for Player 2 and 3, Joker-inspired strategies of global cost-minimal are truly admissible (\autoref{thm:jokeradm}(2)). The inituition is as follows.
From each Joker state, there is a winning and a loosing Player 2 and 3 strategy pair for a given Player 1 action. 
Here we note that global cost-minimal strategies may only suffer from such loosing strategies in Joker states, as they are in control in the other states.
Without memory, Player 2 and 3 fix the choice for every Joker state and Player 1 action of a strategy. In particular, a cyclic move as in \autoref{fig:memstrat} is taken forever.
This leads to the following two conclusions:
\begin{itemize}
    \item Some Joker-inspired global cost-minimal Player 1 strategies have the same winning pairs of Player 2 and 3 strategies, i.e., in particular all strategies that move through the same Joker states.
\item For the other Joker-inspired global cost-minimal Player 1 strategies, we have for each two  Player 1 strategies $\sigma$ and $\sigma'$, a pair of Player 2 and 3 strategies $(\sigma_2,\sigma_3)$, such that $\sigma$ wins from $\sigma_2$ and $\sigma_3$, and $\sigma'$ not, \emph{and vice versa}. In particular, this property holds for strategies that move through different Joker states, and for some strategies (depending on the game topology), that use different actions in the same Joker states (because Player 3 can inspect the Player 1 action played in a state).
\end{itemize}
In either case, the Joker-inspired global cost-minimal Player 1 strategies do not dominate each other (and strategies that are not cost-minimal will not perform better). Consequently, Joker-inspired cost-minimal strategies are admissible against positional playing Player 2 and \nolinebreak 3.

\begin{restatable}{thm}{jokeradm}
\label{thm:jokeradm}
\text{ }
\begin{enumerate}
    \item A Joker-inspired strategy of a global cost-minimal strategy is not always admissible.
    \item A Joker-inspired strategy of a global cost-minimal strategy is admissible for positional Player 2 and 3 strategies.
\end{enumerate}

\end{restatable}

\prooflink{proofthm7}

\section{Joker strategies using fewer moves}
\label{sec:short}

Although multiple Joker attractor strategies can be constructed via \autoref{def:jokerstrat},
their number of moves may not be minimal in the Joker game.
The cause for this is that Joker attractor strategies prefer reaching a Joker state in few moves and using more moves afterwards, over reaching a Joker state in more moves, but using less moves afterwards.
\autoref{fig:costminnojokerstrat} shows that this is not always beneficial for the total number of moves taken towards the goal: Joker attractor strategies may need more moves in total than other cost-minimal strategies.

To address this, we define the \emph{Joker distance attractor} in 
\autoref{def:dattr}. It optimizes for the minimum number of Jokers, as first objective, and prefers fewer moves, as a second objective. The derived strategies (\autoref{def:diststrat}) will use the minimum number of Jokers, like ordinary Joker strategies, but aim to use fewer moves from initial state to goal state.

The Joker distance attractor re-uses the structure of the ordinary Joker attractor sets in its computation. 
To ensure usage of the minimum number of Jokers, only moves within a Joker attractor set, and moves from a Joker state to the next Joker attractor set are considered. Within the subgraph of this restricted set of moves, the total distance from initial state to goal state is minimized. % to compute a distance attractor.
Specifically, the computation proceeds as follows. Goal states $\finally$ are the states $\dattrk{0}$ with distance 0. Then this set is expanded iteratively via the Joker distance predecessor $\dpre$ to $\dattrk{k}$. We use this predecessor to select a state $q$ that has a move to a state $q' \in \dattrk{k-1}$. Only states $q$ that satisfy either of the following conditions (1) or (2) are selected:
\begin{enumerate}
    \item State $q$ is a Joker state, and:
    \begin{itemize}
        \item There is a Joker move from $q$ to $q'$, and
        \item State $q$ is a state of the next Joker attractor set, i.e. $\jrankg(q) = \jrankg(q') + 1$
    \end{itemize}
    \item State $q$ is a non-Joker state of the Joker attractor, and
    \begin{itemize}
        \item There is a move from $q$ to some state $q'\in \dattrk{k-1}$, against any opponent, i.e., a controllable Player 1 action makes the game move from $q$ to $q'$.
        \item State $q$ is state of the same Joker attractor set as $q'$, i.e., $\jrankg(q) = \jrankg(q')$
    \end{itemize}
\end{enumerate}

\begin{defi} \label{def:dpre}
   Let $Q' \subseteq Q$ be a set of states. The \emph{Joker distance predecessor} $\dpre(Q')$ of $Q'$ is: 
    \begin{align*}
        \dpre(Q') &= \{ q \in \jstate \mid \exists a \in \Gamma_1(q), \exists x \in \Gamma_2(q), \exists q' \in \moves(q,a,x):\\
        &\qquad\qquad\qquad\qquad\qquad\qquad\qquad q' \in Q' \wedge \jrankg(q) = \jrankg(q') + 1\} \\
        &\quad \cup \{q \in \jokerg\setminus\jstate \mid  \exists a \in \Gamma_1(q), \forall x \in \Gamma_2(q):\\
        &\qquad\qquad\qquad\qquad\qquad\quad \moves(q,a,x) \subseteq Q' \wedge \jrankg(q) = \jrankg(q')\}
    \end{align*}
\end{defi}
\begin{defi} \label{def:dattr}
    The Player 1 \emph{Joker distance attractor} is $\dattrg$, where:
    \begin{align*}
        \dattrk{0} =&\ \finally\\
        \dattrk{k+1} =&\ \dattrk{k} \cup \dpre(\dattrk{k}\\
        \dattrg =&\ \bigcup_{k \in \mathbb{N}} \dattrk{k}
    \end{align*}
    The function $\djrankg: Q \rightarrow \mathbb{N}$ associates to each state $q \in Q$ its \emph{Joker distance}  $\djrankg(q) = \min\{k \in \mathbb{N} \mid q \in \dattrk{k}\}$.
\end{defi}

\begin{exa}
We compute the Joker distances for states of Joker game of \autoref{fig:costminnojokerstrat}. Its Joker attractor computation is as follows:
\begin{align*}
    \jokerop^0(\gamedist, \smiley)  &= \{5,6,\smiley\}\\
    \jokerop^1_\jokersymb(\gamedist,\smiley) &= \{3,4,5,6,\smiley\}\\
    \jokerop^1(\gamedist,\smiley) &= \{1,2,3,4,5,6,\smiley\}
\end{align*}
The Joker distance attractor is then computed as follows:
\begin{align*}
    \dattrop^0(\gamedist,\smiley) &= \{\smiley\}\\
    \dattrop^1(\gamedist,\smiley) &= \{4,6,\smiley\}\\
    \dattrop^2(\gamedist,\smiley) &= \{2,4,5,6,\smiley\}\\
    \dattrop^3(\gamedist,\smiley) &= \{1,2,3,4,5,6,\smiley\}
\end{align*}
\end{exa}

\begin{figure}[b!]
  \centering
     \begin{tikzpicture}[shorten >=1pt,node distance=1.1cm,>=stealth']
     	\tikzstyle{every state}=[draw=black,text=black,inner sep=1pt,minimum size=11pt,initial text=]
 	\node[state,fill=lightgray,initial,initial where=above] (1) {1};
 	\node[state,fill=lightgray] (2) [below of=1] {2};
  \node[circle,inner sep=-2pt,fill=lightgray] (3) [right of=1] {\LARGE\frownie};
 	\node[circle,inner sep=-2pt,fill=lightgray] (4) [right of=2] {\LARGE\smiley};
 	\path[->]
 	(1) edge [dashed, bend left] node [left] {$a$ $x$} (2)
    (1) edge [dashed, bend right=70] node [left] {$a$ $y$} (2)
    (2) edge [dashed] node [below]  {$a$ $x$} (4)
 	(1) edge node [above] {$b$ $x$} (3)
    (1) edge [dotted] node [right=1mm] {$b$ $y$} (4)
    (3) edge [loop right] node [right] {$a$ $x$} (3)
    (4) edge [loop right] node [right] {$a$ $x$} (4)
 	;
    \end{tikzpicture}
    \caption{The dashed Joker distance strategy uses 0 Jokers and arrives at $\smiley$ in 2 moves against any opponent strategy. The dotted strategy arrives at $\smiley$ in 1 move, if it plays a Joker in state 1 to prevent Player 2 from playing $x$ and reaching {\frownie}. State 1 is not a Joker state, and Player 1 action $b$ in state 1 is not controllable, so the Joker distance attractor will not add state 1 via this $b$ action with distance 1. Instead it will add state 1 via the $a$ action with distance 2.}
    \label{fig:minmovelooses}
    \end{figure}
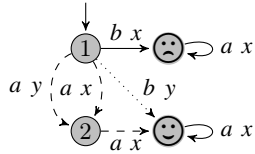

We remark the following about the definition of the Joker distance attractor.
Firstly, we note that the Joker distance attractor has the same structure as the standard attractor (\autoref{def:attr}), except that it uses the special Joker distance predecessor instead of the controlled predecessor of \autoref{def:predecessors}.
Secondly, because the Joker distance predecessor requires that the Joker rank decreases by 1, every time a move through a Joker state is selected, the Joker distance attractor ensures that the minimum number of Jokers is used.
Thirdly, the distance is computed from the number of moves taken, by not preferring moves through non-Joker states over moves through Joker states.
Finally, we note that the computed distance may not be the smallest distance possible. A game may have plays with less moves, that require the use of more Jokers than the minimum. See \autoref{fig:minmovelooses} for an example.
A summary of previous remarks is that the Joker distance attractor is a fixpoint computation (like a standard attractor), that minimizes on (\jrankg, \djrankg), where minimal $\jrankg$ is preferred over minimal $\djrankg$.

Similar to the witnessed attractors of \autoref{def:witness}, we define a witnessed Joker distance attractor (\autoref{def:wdist}), to obtain a corresponding distance-minimizing Joker strategy (\autoref{def:diststrat}). The witnessed Joker distance attractor collects Joker actions (in Joker states) and Player 1 actions (in non-Joker states) that will be composed into a strategy that uses a minimum Jokers and minimum number of moves.

\begin{exa}
As an example, consider game $\gamedist$ of \autoref{fig:costminnojokerstrat} again. We compute the following witnessed Joker distance attractor:
\begin{align*}
    w\dattrop^0(\gamedist,\smiley) &= \emptyset\\
    w\dattrop^1(\gamedist,\smiley) &= \{(4,(a,x,\smiley)),(6,a)\}\\
    w\dattrop^2(\gamedist,\smiley) &= \{(2,a),(4,(a,x,\smiley)),(5,a),(6,a)\}\\
    w\dattrop^3(\gamedist,\smiley) &= \{(1,a),(2,a),(3,(a,x,5)),(4,(a,x,\smiley)), (5,a),(6,a)\}
\end{align*}
In particular, we note that the final set has the move $(1,a)$ to state 2, and not $(1,b)$ to state 3, because $2 \in \dattrop^2(\gamedist,\smiley)$ and $3 \notin \dattrop^2(\gamedist,\smiley)$.
Hence, the corresponding Joker distance strategy chooses action $a$ in state 1, such that {\smiley} is reached in 3 moves.
\end{exa}

\begin{defi} \label{def:wdist}
The \emph{witnessed Joker distance predecessor} $w\dpre(Q')$ of $Q'$ is: 
\begin{align*}
    w\dpre(Q')  & = \{(q,(a,x,q')) \in \jstate \times (\act_1 \times \act_2 \times Q) \mid\\ 
    &\qquad\qquad\qquad q' \in \moves(q,a,x) \cap Q'\wedge \jrankg(q) = \jrankg(q') + 1\}\\
    &\quad \cup \{ (q,a) \in \jokerg\setminus\jstate \times \act_1 \mid \forall x \in \Gamma_2(q):\\
    &\qquad\qquad\qquad\qquad\quad  \moves(q,a,x) \subseteq Q'\wedge \jrankg(q) = \jrankg(q')\}\\
\end{align*}
    The Player 1 \emph{witnessed Joker distance attractor} is $w\dattrg$, where:
    \begin{align*}
  w\dattrk{0} &= \emptyset\\
  w\dattrk{k+1} &= w\dattrk{k}\\
  &\quad \cup \{(q,a) \in w\dpre(\dattrk{k}) \mid q \notin \dattrk{k}\}\\
  w\dattrg &= \bigcup_{k \in \mathbb{N}}w\dattrk{k}\\
    \end{align*}
\end{defi}

  \begin{defi}\label{def:diststrat}
  A strategy $\sigma_1 \in \Sigma_1^\jokersymb(q)$ in $\jokergame$ is a \emph{Joker distance strategy}, if for any $\pi \in \playpref(\jokergame)$ we have: \[\pi^q_{end} \in \dattrg\setminus\finally \implies (\pi^q_{end},\sigma_1(\pi)) \in w\dattrg\]
\end{defi}
 
\begin{figure}[ht!]
\centering
\begin{tikzpicture}[shorten >=1pt,node distance=1.3cm,>=stealth']
\tikzstyle{every state}=[draw=black,text=black,inner sep=1pt,minimum size=11pt,initial text=]
 	\node[state,fill=lightgray,initial,initial where=above] (1) {1};
 	\node[state,fill=lightgray] (2) [right of=1] {2};
    \node[state,fill=lightgray] (3) [left of=1] {3};
    \node[state,fill=lightgray] (4) [below of=3] {4};
 	\node[circle,inner sep=-2pt,fill=lightgray] (5) [below of=1] {\LARGE\smiley};
   \node[circle,inner sep=-2pt,fill=lightgray] (6) [right of=2] {\LARGE\frownie};
 	\path[->]
 	(1) edge [dashed] node [above] {$a$ $y$} (2)
 	(1) edge [dotted] node [above] {$a$ $x$} (3)
    (3) edge [dotted] node [left] {$a$ $x$} (4)
    (4) edge [dotted] node [above] {$a$ $x$} (5)
 	(2) edge [dashed] node [below right] {$a$ $x$} (5)
 	(5) edge [loop right] node [right] {$a$ $x$} (5)
 	(2) edge node [above] {$a$ $y$} (6)
    (6) edge [loop right] node [right] {$a$ $x$} (6)
 	;
 \end{tikzpicture}
 \caption{A cost-minimal strategy with the minimum number of moves is depicted by the dashed edges. It plays a cost-0 $a$ action in state 1, and hopes Player 2 plays $y$ to arrive in state 2 with cost 0. Then it plays a Joker in state 2 to arrive in state $\smiley$ in only 2 moves.
 If Player 2 plays $x$ in state 1, then the game state {\smiley} will be reached with 0 Jokers, in 3 moves. Because state 1 and 2 have the same rank, a Joker distance strategy will use a Joker action from state 1 to  state 3, and 3 moves in total to reach the goal.
 }
 \label{fig:costlessdist}
\end{figure}
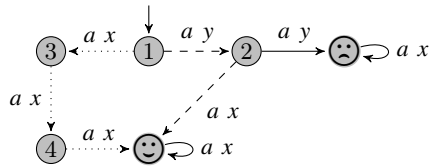

 \autoref{thm:shortcostminimal} states that Joker distance strategies indeed are cost-minimal. Additionally, Joker distance strategies will use a lower or equal number of moves than any Joker attractor strategy. The intuition is that a Joker distance strategy considers more cost-optimal plays than Joker attractor strategies, and then select the plays with the minimum number of moves.
 Against some opponents, however, cost-minimal strategies, that are not Joker attractor strategies, may use less moves than Joker distance strategies. An example is shown in \autoref{fig:costlessdist}.
 Here, the cost-minimal strategy takes a Player 1 move from a Joker state to another Joker state with the same Joker rank. Joker distance strategies only use Joker moves to Joker states, that decrease the Joker rank (see \autoref{def:dpre}), to ensure cost-minimality. It is future work to find a construction for \emph{all} cost-minimal strategies of Joker games, and then optimize this to find strategies with a minimal number of moves.

 \begin{restatable}{thm}{shortcostminimal}
 \label{thm:shortcostminimal}
 Let $\sigma$ be a Joker distance strategy, and $\sigma'$ a Joker attractor strategy, both in Joker game $\jokergame$.
 Then $\sigma$ is cost-minimal (a), and for any play $\pi \in \Outc(\sigma)$ and $\pi' \in \Outc(\sigma')$ we have $\cost(\pi) \le \cost(\pi')$ (b). 
 \end{restatable}

 \prooflink{proofthm8}

\section{Randomized Joker strategies}
\label{sec:randstrats}
\begin{figure}[b!]%[8]
    \centering
    \begin{tikzpicture}[shorten >=1pt,node distance=2cm,>=stealth']
 	\tikzstyle{every state}=[draw=black,text=black,inner sep=1pt,minimum
 	size=11pt,initial text=]
 	\node[state,initial,initial where=above, fill=lightgray] (1) {1};
 	\node[circle,inner sep=-2pt, fill=lightgray] (2) [right of=1] {\LARGE\smiley};
 	\path[->]
 	(1) edge [loop left] node [left] {H T} (1)
 	(1) edge [loop below] node [below] {T H} (1)
 	(1) edge node [above] {H H} (2)
 	(1) edge [bend right] node [below] {T T} (2)
    (2) edge [out=330,in=300,looseness=8] node [right] {H T} (2)
    (2) edge [loop below] node [below] {T T} (2)
    (2) edge [out=50,in=20,looseness=8] node [right] {T H} (2)
    (2) edge [loop above] node [above] {H H} (2)
 	;
    \end{tikzpicture}
    \caption{The penny matching game $\pennygame$: each Player chooses a side of a coin. If they both choose heads (H), or both tails (T), then Player 1 wins, otherwise they play again. Player 1 wins this game with probability 1, by flipping the coin, such that each side has probability $\frac 12$. }
    \label{fig:penny}
\end{figure}
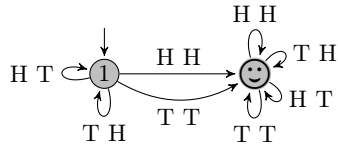

A distinguishing feature of concurrent games is that, unlike in turn-based games, randomized strategies may help to win a game. A classical example is the Penny Matching Game of 
\autoref{fig:penny}.

We show that randomization is not needed for winning Joker games, but does help to reduce the number of Joker moves. However, randomization in Joker states is never needed. 

In \autoref{sec:probattr}, we first set up the required machinery leading to the definition of the probabilistic attractor. We use this in \autoref{sec:probjoker} to define the Joker attractor for randomized strategies: it simply substitutes the standard attractor by the probabilistic attractor. Readers only interested in (novel) results may skip to \autoref{sec:random-joker-properties}: here we revisit the theorems of \autoref{sec:properties-joker-games}, but then for randomized Joker strategies. Finally, in \autoref{sec:nonbenifitrandomjokers} we discuss the (non)benefits of adding randomization to Joker games.

\subsection{Probabilistic attractor}
\label{sec:probattr}

In this section we follow definitions of \cite{dAHK98}, but we adapt them to our concurrent games where $\moves$ returns a set of states.  

In \autoref{def:probstrat}, a randomized strategy for Player 1 or 2 assigns probabilities to a play $\pi$ and next action $a \in \act_1$ or $x \in \act_2$, respectively. A Player 1 strategy $\sigma_p$ for the game $\pennygame$ of \autoref{fig:penny} for example assigns probability $\frac12$ both to $H$ and $T$.
\autoref{def:distr} defines probability distributions and uniform randomness.

\begin{defi}\label{def:distr}
A \emph{(probability) distribution} over a finite set $X$ is a function $\nu:X\to [0,1]$ such that $\sum_{x\in X}\nu(x)=1$. The set of all probability distributions over $X$ is denoted $\Distr(X)$.  Let $\#X$ denote the number of elements of set $X$. A probability function $\nu: X \to [0,1]$ is uniform random for $Y \subseteq X$, if for all $y \in Y$, $\nu(y) = \frac{1}{\# Y}$.
\end{defi}

\begin{defi} \label{def:probstrat}
Let $q \in Q$. A randomized Player $i$ (for $i \in \{1,2\}$) strategy from $q$ is a function $\sigma_i: \pref(q) \rightarrow \Distr(\act_i)$, such that $\sigma_i(\pi)(a)>0$ implies $a \in \Gamma_i(\pi^q_{end})$.
With $\Sigma_i^r(q)$ we denote the randomized strategies for Player $i \in \{1,2\}$ from $q$.
\end{defi}

To define the probability of an outcome, we must not only know how player Players 1 and 2 play, but also how the nondeterminism for choosing from the set of states $\moves(q,a,x)$ is resolved. We introduce Player 3 for resolving this nondeterminism. 
\autoref{def:probplayerthree} defines randomized Player 3 strategies, similar to its non-randomized variant in \autoref{def:playerthree}.
The outcome of a game for a Player 1, 2 and 3 strategy then consists of all plays with actions that have a non-zero probability.

\begin{defi} \label{def:probplayerthree}
  A randomized Player 3 strategy is a function $\sigma_3: \Pi(q) \times \act_1 \times \act_2 \rightarrow \Distr(Q)$, such that $\sigma_3(\pi,a,x)(q')>0$ implies $ q'\in \moves(\pi^q_{end},a,x)$ for all $\pi \in \Pi(q), a \in \Gamma_1(q), x \in \Gamma_2(q)$.
 We write $\Sigma_3^r(q)$ for the set of all randomized Player $3$ strategies from $q$.
 
  The \emph{outcome} $\Outc(\sigma_1,\sigma_2,\sigma_3)$ of randomized strategies $\sigma_1 \in \Sigma_1^r(q)$, $\sigma_2 \in \Sigma_2^r(q)$, and $\sigma_3 \in \Sigma_3^r(q)$ are the plays $\pi \in \Pi^\infty(q)$ such that for all $j \in \mathbb{N}$:
\begin{align*}
\sigma_1(\pi_{0:j})(\pi_{j}^a) > 0 \ \wedge\  \sigma_2(\pi_{0:j})(\pi_{j}^x) > 0 \ \wedge\  \sigma_3(\pi_{0:j},\sigma_1(\pi_{0:j}),\sigma_2(\pi_{0:j}))( \pi_{j+1}^q) > 0
\end{align*}
\end{defi}

Given randomized strategies for Player 1, 2, and 3, \autoref{def:proboutc} defines the probability of a play prefix of the game. This probability is computed as the multiplication of the probabilities given by the strategies for the play prefix. A Player 1 strategy is then an \emph{almost sure winning strategy} if it can win from any Player 2 and Player 3 strategy, with probability 1. Note that for the randomized strategy $\sigma_p$, that chooses $H$ and $T$ with probability $\frac{1}{2}$ in the game $\pennygame$, Player 2 (and 3) cannot prevent with any strategy to not choose the same side of the coin, since Player 1 chooses at random. Since Player 1 may always try again, eventually, always state $\smiley$ will be reached (i.e. with probability 1). Consequently $\sigma_p$ is an almost sure winning strategy.

\begin{defi}
\label{def:proboutc}
Let $\sigma_1 \in \Sigma_1^r(q_0)$, $\sigma_2 \in \Sigma_2^r(q_0)$, and $\sigma_3 \in \Sigma_3^r(q_0)$ be randomized strategies, and let
 $\pi = q_0\langle a_0,x_0 \rangle q_1 \dots \langle a_{j-1},x_{j-1} \rangle q_j$ be a finite play prefix. 
 We define its probability as:
 \vspace{-2mm}
 \begin{align*}
     P(\pi) = \prod_{i=0}^{j-1} \sigma_1(\pi_{0:i})(a_i)\cdot \sigma_2(\pi_{0:i})(x_i)\cdot 
     \sigma_3\left (\pi_{0:i},\sigma_1(\pi_{0:i}),\sigma_2(\pi_{0:i})\right)(q_{i+1})
 \end{align*}
\vspace{-2mm}
 
The strategies $\sigmabar = (\sigma_1, \sigma_2, \sigma_3)$ define a probability space
$(\Omega, \mathcal{F}, \mathcal{P}^{\sigmabar})$ over the set of outcomes. 
A Player 1 strategy $\sigma_1 \in \Sigma_1^r(q)$ is {\em almost sure winning} for reachability goal $R$ if for all $\sigma_2 \in \Sigma_2^r(q)$, and $\sigma_3 \in \Sigma_3^r(q)$,
%we have $\mathcal{P}^{\sigmabar}[\{\pi\in \Pi(q) | \pi \text{ reaches } R\}] > 0$.
we have: 
$\mathcal{P}^{\sigmabar}[\{\pi\in \Outc(\sigma_1,\sigma_2,\sigma_3) \mid \pi \text{ is winning}\} ]=1$. Strategy $\sigma_1$ is \emph{sure winning} if for all $\sigma_2 \in \Sigma_2^r(q)$, and $\sigma_3 \in \Sigma_3^r(q)$, and for any $\pi \in \Outc(\sigma_1,\sigma_2,\sigma_3)$, $\pi$ is winning.
\end{defi}

\autoref{def:pattr}
restates the definition of the probabilistic attractor \cite{dAHK98}, for our concurrent games. The Player 1 probabilistic attractor $\pattrf$ is computed iteratively to find the largest set of states that Player 1 can confine the game to. To ensure this confinement,  Player 1 restricts the set of actions she chooses from, for each state. These Player 1 actions for states correspond to the witness $\cpre_1^w$ as defined in \autoref{def:witness}. This witness gets refined in each iteration $\pattrfi{k}$, by computing the states Player 1 and 2 can confine the game to for the current witness, namely the set of states $\sattr_i(Q')$ (\autoref{def:safe}). As noted in \cite{dAHK98}, efficient data structures can be used for efficient computation \cite{Beeri80}. In \autoref{def:safe} we give the (simple) naive fixpoint computation.

\begin{exa}\label{exmp:pattr}
As an example, consider the game $\pennygame$ of \autoref{fig:penny} again. The computation of the probabilistic attractor is then as follows:
\begin{align*}
p\attrop^0_1(\pennygame,\{\smiley\}) &= \{1,\smiley\}\\
p\attrop^0_2(\pennygame,\{\smiley\}) &= \sattr_2(p\attrop^0_1(\pennygame,\{\smiley\})\setminus \{\smiley\},\cpre^w_1(p\attrop^0_1(\pennygame,\{\smiley\})))\\
&= \sattr_2(\{1,\smiley\}\setminus\{\smiley\},\cpre(\{1,\smiley\}))\\
&= \sattr_2(\{1\},\{(1,H),(1,T),(2,H),(2,T)\})\\
&= \emptyset \text{ (see below)}\\
\sattr_2^0(\{1\},\{(1,H)&,(1,T),(2,H),(2,T)\}) = \{1\}\\
\sattr_2^1(\{1\},\{(1,H)&,(1,T),(2,H),(2,T)\}) = \{1\} \cap p\cpre_2(\{1\}) = \{1\}\cap \emptyset = \emptyset\\
&\phantom{,(1,T),(2,H),(2,T)\}) }=\sattr_2^2(\{1\},\{(1,H),(1,T),(2,H),(2,T)\})\\
p\attrop^1_1(\pennygame,\{\smiley\}) &=\sattr_1(p\attrop_1^0(\pennygame,\{\smiley\})\setminus p\attrop_2^0(\pennygame,\{\smiley\}),\cpre_1^w(p\attrop_1^0(\pennygame,\{\smiley\})))\\
&=\sattr_1(\{1,\smiley\},\{(1,H),(1,T),(2,H),(2,T)\}) = \{1,\smiley\} \text{ (see below)}\\
\sattr_1^0(\{1,\smiley\},\{(1,H)&,(1,T),(2,H),(2,T)\}) = \{1,\smiley\}\\
\sattr_1^1(\{1,\smiley\},\{(1,H)&,(1,T),(2,H),(2,T)\}) = \{1,\smiley\} \cap p\cpre_1(\{1,\smiley\}) = \{1,\smiley\}
\end{align*}
\end{exa}

\begin{defi}\label{def:safe} The \emph{probabilistic controlled predecessor} $p\cpre_i(Q',W)$, and \emph{confinement sets} $\sattr_i$ for Player $i \in \{1,2\}$ are defined as follows:
     \begin{align*}
  &p\cpre_1(Q',W) = \{\ q \in Q \mid \exists a\in\Gamma_1(q):  (q,a) \in W \wedge \forall x \in \Gamma_2(q): \moves(q,a,x) \subseteq Q'\}\\
  &p\cpre_2(Q',W) = \{\ q \in Q \mid \exists x\in\Gamma_2(q), \forall a \in \Gamma_1(q): (q,a) \in W \implies \moves(q,a,x) \subseteq Q'\}\\
  &\sattr^{0}_i(Q',W) = Q'\\
  &\sattr^{k+1}_i(Q',W) = Q'\ \cap p\cpre_i(\sattr_i^k(Q',W),W)\\
  &\sattr_i(Q',W) = \bigcap_{k \in \mathbb{N}} \sattr^{k}_i(Q',W)
  \end{align*}
\end{defi}

 \begin{defi} \label{def:pattr} The Player 1 probabilistic attractor is $\pattrf$, where:
  \begin{align*}
      &\pattrfi{0} = Q\\
      &\pattrsi{k} = \sattr_2(\pattrfi{k} \setminus R, \cpre_1^w(\pattrfi{k}))\\
      &\pattrfi{k+1} = \sattr_1(\pattrfi{k} \setminus \pattrsi{k}, \cpre_1^w(\pattrfi{k}))\\
      &\pattrf = \bigcap_{k \in \mathbb{N}} \pattrfi{k}
  \end{align*}
The function $\parankg: Q \times 2^Q \rightarrow \mathbb{N}$ associates to each state $q\in Q$ a rank $\parankg(q, R) = \min \{k \in \mathbb{N} \mid q \in \pattrfi{k}\}$.
\end{defi}

A randomized Joker strategy $\sigma \in \Sigma_1^r(q)$ is defined from the witness computed in the last iteration of the probabilistic attractor: the strategy chooses a Player 1 action $a$ from state $q$ uniformly at random for any $(q,a) \in \cpre_1^w(\pattrf\setminus R)$. 
From the computation of Example~\ref{exmp:pattr}, we have that $(1,H),(1,T) \in \cpre^w_1(\pattrf)$. Hence, the strategy $\sigma_p$ for game $\pennygame$ is a probabilistic attractor strategy, since it assigns probability $\frac{1}{2}$ to $H$ and $T$ each from state 1.%, for plays ending in state 1. That $H$ is chosen with probability 1 in state {\smiley} is irrelevant, since {\smiley} is a goal state.

\begin{defi} \label{def:pstrat}
A randomized Joker strategy is a strategy that is uniform random for $\cpre_1^w(\pattrf)$, i.e. for any $\pi \in \Pi(\game)$ such that $\pi_{end}^q \in \pattrf\setminus R$ we have $\sigma(\pi)(a) = \dfrac{1}{\#\{(\pi^q_{end},a) \in \cpre_1^w(\pattrf\setminus R)\}}$.
\end{defi}

\subsection{Probabilistic Joker attractor}
\label{sec:probjoker}

In \autoref{sec:properties-joker-games}, several fundamental properties on Joker attractor strategies were given. In this subsection, we will see that these theorems hold for randomized strategies too.
In particular, we consider randomized Player 2 and 3 strategies, against randomized Joker strategies. We construct the latter strategies by adapting the Joker attractor (\autoref{def:jokers}): we replace the standard attractor $\attrf$ by the probabilistic attractor $\pattrf$. The result is a probabilistic Joker attractor (\autoref{def:pjokers}). The corresponding randomized Joker strategies (\autoref{def:pjokerstrat}) use a Joker action in a Joker state, as usual, but use randomization, like randomized strategies, in other states.

\begin{defi} \label{def:pjokers}
 The Player 1 probablistic Joker attractor is $\pjokerg$, where:
 \begin{align*}
  \pjoker{0} &= \pattrf\\
  \pjokerreal{k+1} &= \pjoker{k}\ \cup  \pre(\pjoker{k})\\
  \pjoker{k+1} &= \pattrop(\pjokerreal{k+1})\\
  \pjokerg &= \bigcup_{k \in \mathbb{N}} \pjoker{k}
  \end{align*}
 We call $\pjokerg$ the probabilistic Joker attractor of $\game$. The probabilistic Joker states are 
  $\pjstate = \bigcup_{k \in \mathbb{N}} \pjokerreal{k}\setminus\pjoker{k}$.
 To each probabilistic Joker attractor $\pjokerg$ we associate a probabilistic Joker rank function $\pjrankg : Q \rightarrow \mathbb{N}$, where for each state $q \in Q$ we define $\pjrankg(q) = \min \{k \in \mathbb{N} \mid q \in \pjoker{k}\}$.
\end{defi}

\begin{figure}[t!]%[8]
    \centering
    \begin{tikzpicture}[shorten >=1pt,node distance=2cm,>=stealth']
 	\tikzstyle{every state}=[draw=black,text=black,inner sep=1pt,minimum
 	size=11pt,initial text=]
 	\node[state, fill=lightgray] (1) {1};
 	\node[circle,inner sep=-2pt, fill=lightgray] (2) [right of=1] {\LARGE\smiley};
    \node[state,initial,initial where=above, fill=lightgray] (0) [left of=1] {0};
    \node[circle,inner sep=-2pt, fill=lightgray] (3) [left of=0] {\LARGE\frownie};
 	\path[->]
 	(1) edge [loop above] node [above] {H T} (1)
 	(1) edge [loop below] node [below] {T H} (1)
 	(1) edge node [above] {H H} (2)
 	(1) edge [bend right] node [below] {T T} (2)
    (2) edge [out=330,in=300,looseness=8] node [right] {H T} (2)
    (2) edge [loop below] node [below] {T T} (2)
    (2) edge [out=50,in=20,looseness=8] node [right] {T H} (2)
    (2) edge [loop above] node [above] {H H} (2)
    (0) edge [bend left] node [above] {H H} (1)
    (0) edge [bend right] node [below] {T H} (1)
    (0) edge [bend left] node [below] {H T} (3)
    (0) edge [bend right] node [above] {T T} (3)
    (3) edge [out=150,in=120,looseness=8] node [left] {T H} (3)
    (3) edge [loop below] node [below] {T T} (3)
    (3) edge [out=230,in=200,looseness=8] node [left] {H T} (3)
    (3) edge [loop above] node [above] {H H} (3)
 	;
    \end{tikzpicture}
    \caption{The penny matching game extended with two additional states. A randomized Joker strategy uses either of Joker actions (0,H,H) and/or (0,T,H) with probability 1 from state 0, and actions H and T with probability $\frac{1}{2}$ each from state 1. A non-randomized Joker strategy would use Jokers both in state 0 and 1.}
    \label{fig:pennyext}
\end{figure}
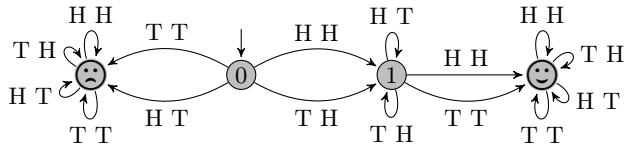

\begin{exa}
Consider for example the game in \autoref{fig:pennyext}. Here we added states 0 and {\frownie} to the penny matching game of \autoref{fig:penny}, and state 0 is the initial state. From state 0, Player 2 is in full control: if Player 2 chooses H, the game moves to state 1, while if he chooses T, the game moves to {\frownie}, nomatter what Player 1 chooses. In the computation the probabilistic Joker attractor, we then find that $\{1,\smiley\} \subseteq \pattrf$, $0 \in \pjokerreal{1}$, and that $\frownie \notin \pjokerg$.
\end{exa}

For randomized Joker strategies (\autoref{def:pjokerstrat}), we first need to define the probabilistic version of the witnessed Joker attractor (\autoref{def:pwitness}).
The difference with the witnesses of \autoref{def:witness} is that we extract the witness of a probabilistic attractor, by directly applying $\cpre^w_1$ on this set of states, exactly as the definition of the randomized Joker strategy (\autoref{def:pstrat}).
\autoref{def:pjokerstrat} defines that in Joker states, a randomized Joker strategy uses the witness (i.e. the Joker action) of the probabilistic Joker attractor, and in other states the strategy uses the witness (i.e. controlled predecessor) of the probabilistic Joker attractor.

 \begin{defi} \label{def:pwitness}
We define the \emph{witnessed} probabilistic Joker attractor $wp\jokerg$ as follows:
  \begin{align*}
  w\pjoker{0}&= \emptyset\\
  w\pjokerreal{k+1} &= w\pjoker{k} \cup \{ (q,a) \in w\pre(\pjoker{k}) \mid q \notin \pjoker{k} \}\\
   w\pjoker{k+1} &=  \cpre^w_1(\pattrop(\game,w\pjokerreal{k+1}))\\
  w\pjokerg &= \bigcup_{k \in \mathbb{N}}w\pjoker{k}
  \end{align*}
  \end{defi}

  \begin{defi}\label{def:pjokerstrat}
  A randomized strategy $\sigma_1 \in \Sigma_1^{\jokersymb,r}(q)$ in $\jokergame$ is a \emph{randomized Joker  strategy}, if for any $\pi \in \playpref(\jokergame)$ with $\pi^q_{end} \in \pjokerg$ and any $a \in \Gamma_1(\pi^q_{end})$ we have:
  \begin{itemize}
      \item  $ \sigma_1(\pi)(a) > 0 \implies (\pi^q_{end}, a) \in  w\pjokerg$,
      \item $\pi^q_{end} \in \jstate \implies a \in \Gamma_1^\jokersymbreal(\pi^q_{end}) \wedge \sigma_1(\pi)(a) = 1$, and
      \item $\pi^q_{end} \notin \jstate \implies a \in \Gamma_1(\pi^q_{end})$, and $\sigma_1$ is uniform random in $\pi^q_{end}$ for its non-zero probability actions $a$
  \end{itemize}
\end{defi}

The extended penny matching game of \autoref{fig:pennyext} shows the advantage of randomized Joker attractor strategies over non-randomized Joker attractor strategies: randomization may help to use less Jokers (\autoref{thm:probjokercostless}).

\begin{restatable}{thm}{probjokercostless}
\label{thm:probjokercostless}
    Let $\sigma$ be a randomized Joker strategy, and $\sigma'$ a Joker attractor strategy. Then $\cost(\sigma) \le \cost(\sigma')$.
\end{restatable}

\prooflink{proofthm9}

\subsection{Properties of Joker games with randomization}
\label{sec:random-joker-properties}

In this subsection, we revisit the theorems of \autoref{sec:properties-joker-games} for randomized strategies. \autoref{thm:probjokerwinreach} states that all outcomes of a \emph{randomized} Joker strategy use exactly the same number of Jokers. The intuition for this is that the probabilistic Joker attractor $\pjokerg$ and the Joker attractor $\jokerg$ both use the predecessor $\pre$ in the construction for the Joker action. The difference is the use of $\pattrf$ versus $\attrf$, i.e. in the penny matching game of \autoref{fig:penny}, Joker strategies use 1 Joker to move from state 1 to {\smiley}, while randomized Joker strategies use 0 Jokers.

\begin{restatable}{thm}{probnrjokermoves}
\label{thm:probnrjokermoves}
Let $q \in \pjokerg$, and let $\jokergame$ be the Joker game, where all players may use randomized strategies.
Then: 
\begin{enumerate}
    \item Let $\sigma^J_1 \in \Sigma_1^{\jokersymb,r}(q)$ be a randomized Joker strategy in $\jokergame$.
    Then any play $\pi \in \Outc(\sigma^J_1)$ has \emph{exactly} $\pjrankg(q)$ Joker actions in winning prefix $\pi_{0:\ind(\pi)}$.
    \item Let $\sigma_1 \in \Sigma_1^{\jokersymb(q),r}$ be a randomized, cost-minimal strategy in $\jokergame$.
    Then any play $\pi \in \Outc(\sigma_1)$ has \emph{at most} $\pjrankg(q)$ Joker actions in the winning prefix $\pi_{0:\ind(\pi)}$.
\end{enumerate}
\end{restatable}
\prooflink{proofthm10}\bigskip

In \autoref{thm:probjokerwinreach} we state that in (Joker) games with randomized strategies, the probabilistic Joker attractor coincides with the states from where a goal state can be reached in the concurrent game, and with the states from where Player 1 can win in the Joker game. The intuition for the proof is that Joker attractor strategies may replace the power of randomization (of randomized Joker strategies) by using additional Jokers actions (see also \autoref{sec:nonbenifitrandomjokers}).

\begin{restatable}{thm}{probjokerwinreach}
\label{thm:probjokerwinreach}
Consider the Joker games $\jokergame$, where all players may use randomized strategies.
 Let $\reach = 
 \{q \in Q \mid q \text{ can reach a state } q'\in R\}$.
 Then
 \begin{align*}
     \winrop(\jokergame,R) = \pjokerg = \reach
 \end{align*}
\end{restatable}

\prooflink{proofthm11}\bigskip

\autoref{thm:probjokerstratwinning} states that randomized Joker strategies are cost-minimal.
The intuition is that the probabilistic Joker attractor has the same structure as the Joker attractor. 

\begin{restatable}{thm}{probjokerstratwinning}
 \label{thm:probjokerstratwinning}
 Consider the Joker games, where all players may use randomized strategies.
     Then any randomized Joker strategy is cost-minimal.
\end{restatable}

\prooflink{proofthm12}\bigskip

Similar to determinacy for Joker games without randomized strategies (\autoref{thm:jokerdetermined}), we have that determinacy of Joker games \emph{with} randomized strategies (\autoref{thm:probjokerdetermined}) follows from \autoref{thm:probjokerwinreach}.

\begin{restatable}{thm}{probjokerdetermined}
\label{thm:probjokerdetermined}
Consider the Joker games, where all players may use randomized strategies.
Then these Joker games are determined.
\end{restatable}

\prooflink{proofthm13}

\subsection{The (non-)benefits of randomization in Joker games}
\label{sec:nonbenifitrandomjokers}

We show that Joker games do not need randomized strategies: 
if Player 1 can win a Joker game with a randomized strategy, then she can win this game with a non-randomized strategy.
This result is less surprising than it may seem (\autoref{thm:jokerrandom}(1)), since Joker actions are very powerful:
they can determine the next state of the game to be any state reachable in one move.
In the penny matching game of \autoref{fig:penny}, Player 1 may in state 1 just take the Joker move $(H,H,\smiley)$ and reach the state {\smiley} immediately. With randomization, Player 1 can win this game without using Joker moves.

We note however that we only need to use the power of Jokers in Joker states (\autoref{thm:jokerrandom}(2)). We can use the probabilistic Joker attractor (\autoref{def:pjokers}) to attract to Joker states with probability 1, and then use a Joker move in this Joker state, where even using randomization, the game cannot be won with probability 1. The Jokers used in these Joker states then only needs to be played deterministically, i.e. with probability 1 (\autoref{thm:jokerrandom}(3)). The intuition for this last statement is that a Joker move determines the next state completely, so by choosing the `best' next state there is no need to include a chance for reaching any other state. In \autoref{thm:probjokerstrat} we note that, in particular, the randomized Joker strategies of \autoref{def:pjokerstrat} satisfy this last property \autoref{thm:jokerrandom}(3).

\begin{restatable}{thm}{jokernotrandom}
\label{thm:jokerrandom}
If a state $q\in Q$ of Joker game $\jokergame$ has an almost sure winning strategy $\sigma_1^r \in \Sigma_1^{r,\jokersymb}(q)$, then
\begin{enumerate}
\item  she also has a winning non-randomized strategy.
\item  she also has an almost sure winning strategy that only uses Jokers in Joker states.
\item she also has an almost sure winning strategy that only uses Jokers in Joker states, such that these Jokers can all be played with probability 1.
\end{enumerate}
\end{restatable}

\prooflink{proofthm14}

\begin{restatable}{thm}{probjokerstrat}
\label{thm:probjokerstrat}
Any strategy from \autoref{def:pjokerstrat} satisfies \autoref{thm:jokerrandom}(3).
\end{restatable}

\prooflink{proofthm15}

\section{Experiments}
\label{sec:experiments}

We illustrate the application of Joker games in model-based testing (MBT).

\paragraph{MBT}
In MBT, the model specifies the desired interaction of the system-under-test (SUT) with its environment, e.g. the user or some other system. The tester (i.e. this is a special user) interacts with the SUT to find bugs, i.e. undesired interactions, of the system. A test case specifies the inputs the tester provides, given the outputs that the SUT provides in the interaction. In MBT, test cases are derived from the model. By executing the test cases on the SUT, it can be checked whether the SUT shows the desired interaction, as specified by the model. 

Because the number of test cases that can be derived from the model is usually infinite, a selection criteria is used to select a finite number. One standard criteria is to derive test cases such that for each state in the model, a test case reaching this state in the model is selected. When the test cases have been executed on the SUT, we say that all states are covered. 

The derivation of test cases that cover all states is non-trivial, since the tester may not have full control over the outputs of the SUT. The SUT may choose its outputs in such a way that prevents reaching some states. On the other hand, most SUTs are not adversarial by nature, but they are also not necessarily cooperative. 

\paragraph{Testing as a game.}
Therefore, we apply our Joker games to find test cases that reach a state with the least help, i.e. the minimum number of Jokers. To do this, we represent the tester as Player 1 and the SUT as Player 2 in our Joker games. Specifically, we translate the model-based testing model to a concurrent game, compute Joker attractor strategies and their Joker-inspired strategies, and translate the latter strategies to test cases. The first and last step are performed via the translations provided by Van den Bos and Stoelinga \cite{mbtgames}.

In the translation from model to game, we use that in each game state,
Player 1/Tester has three options: stop testing, provide one of her inputs to the SUT, or observe an output of the SUT. Player 2/SUT has two options: provide an output to the tester, or do nothing.
Hence Player 1 and 2 decide concurrently what their next action is.
The next state is then determined as follows: if the Tester provides an input, and the SUT does nothing, the input is received and processed by the SUT. If the Tester observes the SUT (virtually also doing no action of its own), and the SUT provides an output, the output was processed by the SUT and sent to the Tester. If the Tester provides an input, while the SUT also provides an output, then several ways of solving this conflict are possible, such as input-eager, output-eager \cite{mbtgames}. We opt for the {\em nondeterministic} solution,  where the picked action is chosen nondeterministically. Note that this corresponds with having a set of states $\moves(q,a,x)$. 

\paragraph{Testing experiments.}
We investigate the effectiveness of Joker strategies in their application in MBT, by comparing the Joker-based testing, as explained above, with randomized testing.
In randomized testing, the tester selects an action (i.e. an input, or doing nothing) uniformly at random, or chooses to stop testing.
In Joker-based testing the test case executes the actions of the Joker-inspired strategy, or decides to stop testing. 
By using Joker-inspired strategies as test cases, we allow for a fair and realistic (no use of Joker actions) comparison.  

\paragraph{Case studies.} We applied our experiments on four case studies from \cite{VandenBosVaandragerJournal}: the opening and closing behaviour of the TCP protocol (26 states, 53 transitions) \cite{tcp}, an elaborate drinks vending machine for drinks (269 states,687 transitions) \cite{vendmach,modelsward17},
the Echo Algorithm for leader election (105 states, 275 transitions) \cite{fokkink2013distributed}, and
the Dropbox functionality for file storage in the cloud (752 states, 1520 transitions) \cite{hughesdropbox,tretmansdropbox}.

\paragraph{Experimental setup.}
For each case study, we randomly selected the different goal states: 5 for the (relatively small) TCP case study  and 15 for the other cases. 
For each of these goals, we extract a Joker attractor strategy.
Specifically, we compute Joker attractor strategies (\autoref{def:jokerstrat}) from the witnessed Joker attractor (\autoref{def:witness}), and take their Joker-inspired strategies (\autoref{def:insp}).
If there were multiple resulting strategies for one goal state, we select one at random.
Then we translate the obtained Joker-inspired strategies to test cases (one for each goal). 
Also, we translate the model-based testing model to a concurrent game. The latter two translations are performed as defined by Van den Bos and Stoelinga \cite{mbtgames}.
We run each of the obtained Joker test cases 10.000 times. Also, we run 10.000 random test cases. A random test case chooses a possible Player 1 action at uniformly random.
All Joker test cases and random test cases choose to stop testing with (the same) probability $p$, and choose an input action with probability $1-p$.

Test case execution is done according to the standard model-against-model testing approach \cite{VandenBosVaandragerJournal}. In this setup, an impartial SUT is simulated from the model, by making the simulation choose any SUT-action uniformly at random, and by resolving any non-determinism by choosing a state from $\moves(q,a,x)$ uniformly at random too.
Hence, when a Joker- or random test case proposes a Player1/Tester action, the simulation chooses a Player 2/SUT-action, and then resolves any non-determinism. When the Joker- or random test case proposes to stop testing, the simulation ends. The finite play, with Player 1 and 2 actions, and the visited states is then recorded as the result of the test case execution. This way, we collect the test results of all 10.000 Joker test cases and all 10.000 random test cases, per goal, of each case study.
Per goal and case study, and for each of the two sets of 10.000 test cases we then compute the following numbers:
(1) the percentage of runs that reach the goal state, and
(2) the average number of moves to reach the goal state, if a goal state was reached.

\begin{figure}
    \centering
    \includegraphics[height=0.3\textheight]{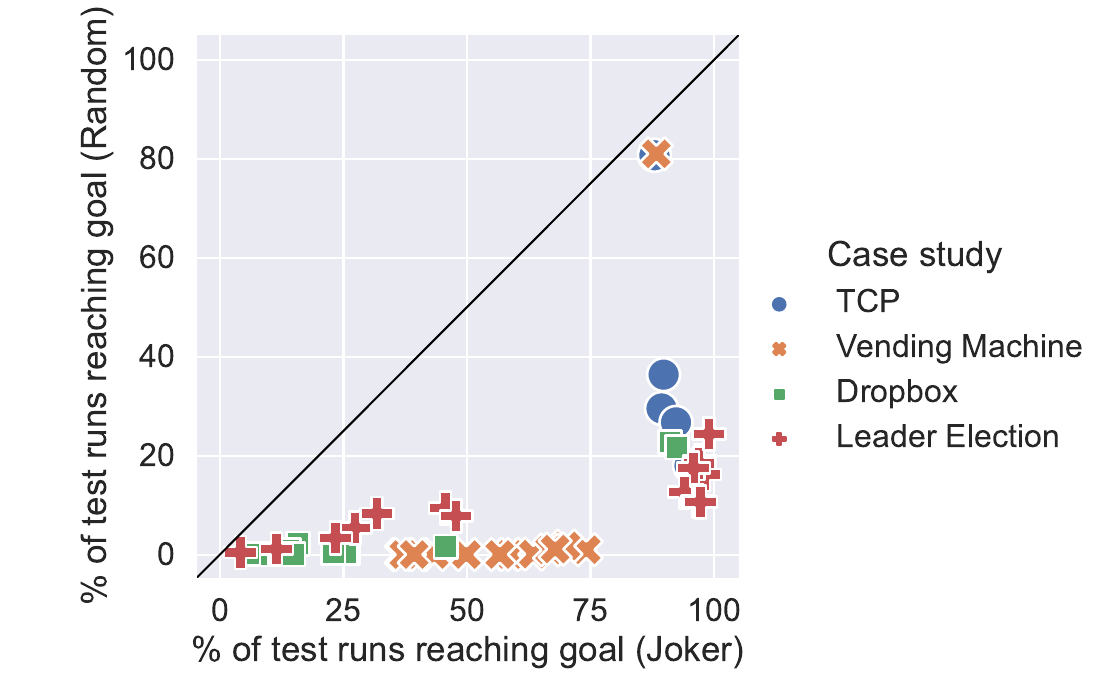}
    \caption{Joker-based versus randomized testing: percentage of tests reaching goal.}
    \label{fig:exp-perc}
\end{figure}

\paragraph{Results.}
\autoref{fig:exp-perc} shows the experimental results for computation (1). In this graph, a point represents one goal state of a case study. The x-coordinate is the computed percentage (1) of the Joker test cases, and the y-coordiante is the computed percentage (1) of the random test cases. This way we compare the percentage of Joker test cases that reach the goal state with the percentage of random test cases that reaches the goal state.
Hence, if a point is positioned below the diagonal, the Joker test cases reached the goal state more often than the random test cases. We see that this is the case for all points. Also, a majority of points is close to 0\% for the random test cases, while Joker test cases are (much) more successful for most goals.

\begin{figure}[ht!]
\centering
 \includegraphics[height=0.29\textheight]{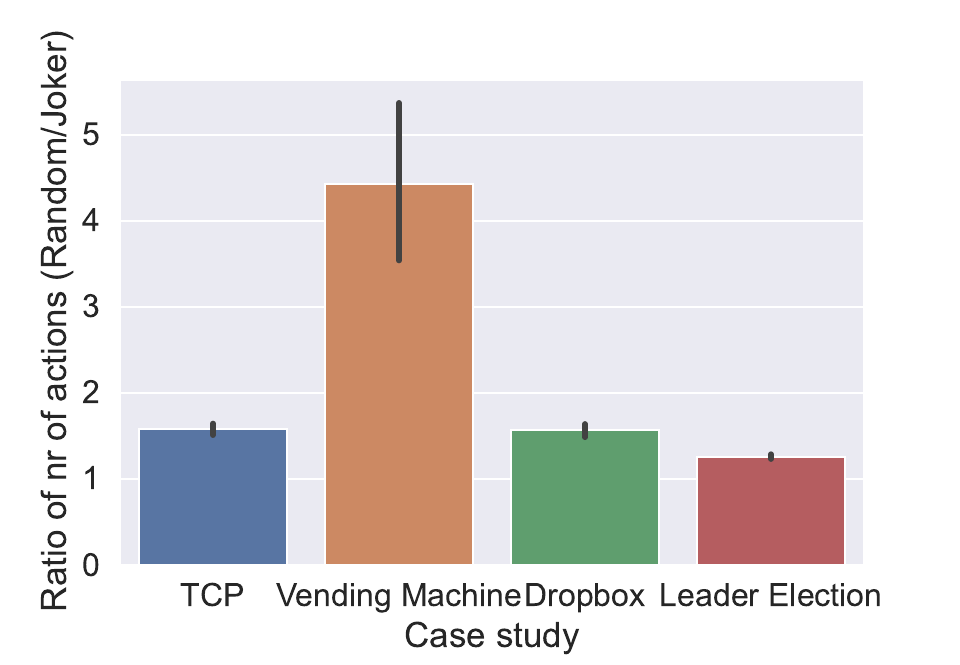}
\caption{Ratio of average number of moves to the goals, where the average number of the random test cases is divided by the average number of the Joker test cases.}
\label{fig:exp-bar}
\end{figure}

\autoref{fig:exp-bar} shows the experimental results for computation (2).
To construct the graph, we computed for each goal, and separately for the sets of Joker test cases and random test cases, the average number of actions for reaching the goal. Hence, for each goal we obtain two average numbers: a number $r$ for the random test cases, and a number $j$ for the Joker test cases. For each goal, we then obtain the ratio $r/j$. If a test case did not reach a goal, the test case is not used to compute the average, so e.g. if $x$ out of the 10.000 test cases reached the goal, the average number of moves is computed from the number of moves of the $x$ test cases.

One bar in \autoref{fig:exp-bar} represents one case study, with all its goals. The black line depicts the variation between the ratios for the goals. Hence, we see that for all case studies but the Vending Machine, the ratios are close to each other. A possible reason for the high variation of ratios for the Vending Machine case study, is that we can see in \autoref{fig:exp-perc} that there are especially few random test cases that reach the goals of the Vending Machine case study, such that the average is computed over the number of moves of those few test cases. Moreover, we omitted results for 5 goal states of the Vending Machine case study, as these states were not reached in any of the 10.000 runs of random testing, while there were $> 3500$ successful Joker runs for each of those goals. In general, for all the case studies, we see that the ratio is above 1, meaning that the Joker test cases use less moves than the random test cases.

The overall conclusion from the graphs of \autoref{fig:exp-perc} and \autoref{fig:exp-bar} is that the Joker test cases clearly outperform the random test cases, both in reaching goal states, and the number of moves needed to reach a goal state.
Our experimental results can be reproduced with the artefact \cite{artefact}.

\section{Conclusions}
\label{sec:concl}

We introduced the notion of Joker games, showed that its attractor-based Joker strategies use a minimum number of Jokers, and proved properties on determinacy, minimization of number of moves, randomization, and admissible strategies in the context of Joker games. In experiments we showed the effectiveness of Joker strategies when applied in model-based testing.

In future work, we would like to extend the experimental evaluation of Joker strategies on applications, and investigate more cost-minimal strategies. For example, we would like to also find the cost-minimal strategies that are currently not found via the Joker attractor construction. Furthermore, we are interested in other multi-objective strategies than our Joker distance strategies, i.e. strategies that are cost-minimal and also optimized for some other goal. Specifically, inspired by \cite{synthesissafety}, we would like to investigate how to select and construct the Joker strategies that are the most  robust against `bad' opponent actions. Such `bad' actions could e.g. move the game to a state where the game cannot be won anymore, or to a state from where many Jokers need to be spend to reach the goal. Lastly, we would like to investigate the relation between our work on Joker games and the work of Annand et al.  \cite{AnandMNS23}, which comprises adequately permissive assumptions on (turn-based) games with  $\omega$-regular winning conditions.

\bibliographystyle{alphaurl}
\bibliography{lib}

\appendix

\clearpage
\appendix
\section*{Appendix}
\label{sec:appendix}

\section{Proofs}

\hypertarget{proofthm1}{
\jokeriscost*
}

\noindent
\textit{Proof \autoref{thm:costgameisjokergameNew}.}
First consider case $q \notin \jokerg$. By \autoref{def:jokers} we then have that $\jrankg(q) = \infty$.
By \autoref{thm:jokerwinreach} we have $q \notin \reach$.
Hence $q$ cannot reach goal $R$ with any play. By \autoref{def:costfunction}, we then have that any play has cost $\infty$, so any strategy has cost $\infty$, and then $\cost(q) = \infty.$
Hence, $\jrankg(q) = \cost(q)$.

Now consider  $q \in \jokerg$, so $q \in \reach$. By \autoref{def:jokerstrat} we then have a Joker attractor strategy. By \autoref{thm:jokerstratwinning}, this strategy is cost-minimal. By \autoref{thm:nrjokermoves} it uses $\jrankg(q)$ Jokers in any play, so its cost is $\jrankg(q)$.
Since $\cost(q)$ is defined as the cost of cost-minimal strategies (\autoref{def:costfunction}), we have that $\jrankg(q) = \cost(q)$. \qed

\hypertarget{proofthm2}{
\nrjokermoves*
}

\noindent
\textit{Proof \autoref{thm:nrjokermoves}.}

\noindent
(1) 
We first prove lemma (W):  $\sigma_1^J$ is a winning strategy in Joker game $\jokergame$.
Let $q'$ be a state in a play of $\Outc(\sigma_1^J)$.  Observe the following three cases:
\begin{itemize}
    \item If $q' \in \attrf$, then $\sigma_1^J$ is a classical attractor strategy from $q$ that is winning \cite{dAHK98}, so reaches $R$, without using any Joker actions.
    \item If $q' \in \joker{k}\setminus\jstate$, $q'$ is a non-Joker state of the $k$-th Joker attractor set, such that $\sigma_1^J$ is a classical attractor strategy from $q'$ with goal $\jokerreal{k}$. Hence, $\sigma_1^J$ will surely reach one of the Joker states $\jokerreal{k}$, without using any Joker actions.
    \item If $q' \in \jstate$, then by the definition of a Joker attractor strategy (\autoref{def:jokerstrat}), $\sigma_1^J$ uses a Joker action derived from the predecessor $\pre$, such that the Joker action causes the game to reach state $q'' \in \joker{k}$ from $q' \notin \joker{k}$. The use of $\pre_w$ in the definition of a Joker attractor strategy (\autoref{def:jokerstrat}) exactly corresponds with the use of $\pre$ in the definition of the Joker attractor (\autoref{def:jokers}), so we have $\jrankg(q'') + 1 = \jrankg(q')$, and spend 1 Joker action making this move.
\end{itemize}
Consequently, $\sigma_1^J$ is a winning strategy, so (W) holds.
Moreover, using 1 Joker action, reduces the rank of the next state by 1. Using the attractor in non-Joker states, costs no Joker actions. Since each of the $\jrankg(q)$ encountered Joker states we use 1 Joker action, we spend exactly $\jrankg(q)$ Joker actions in total.

\noindent
2) By \autoref{thm:jokerstratwinning} we have that a Joker attractor strategy is cost-minimal, so any other cost-minimal strategy must have cost $\jrankg(q)$. That the cost can be less than $\jrankg(q)$ for some plays follows from \autoref{fig:costless}. \qed

\hypertarget{proofthm3}{
 \jokerwinreach*
}

\textit{Proof \autoref{thm:jokerwinreach}.}
To prove $\reach \subseteq \winrop(\jokergame,\finally)$, we observe that a state $q' \in R$ that can be reached from a state $q$ has a play $q_0 \langle a_0, x_0 \rangle q_1 \dots q_n$ with $q_0 = q$ and $q_n = q'$. This play allows constructing the winning strategy $\sigma \in \Sigma_1^\jokersymb(q)$ in the Joker game, that only plays Joker actions to imitate the play: $\sigma(q_0 \dots q_k) = (a_k,x_k,q_{k+1})$ for all $0 \le k < n$ in $\jokergame$.

We conclude $\winrop(\jokergame,\finally) \subseteq \reach$  from the three facts that (1)
a state with a winning strategy in $\jokergame$ has a play reaching $R$, and (2) $\jokergame$ has the same states as $\game$, and (3) playing a Joker action in $\jokergame$ corresponds to a regular move in $\game$.

From lemma (W) from the proof of \autoref{thm:nrjokermoves}(1) we can derive that any state in $\jokerg$ has a winning strategy in $\jokergame$, so $q \in \winrop(\jokergame,\finally)$. Consequently we have: $\jokerg \subseteq \winrop(\jokergame,\finally)$.

Lastly, to prove that $\reach \subseteq \jokerg$, we observe that $q \in (\pre)^k(\jokerg)$ $\iff q \in \jokerg$ for any $k \in \mathbb{N}$ by the definition of $\jokerg$ (\autoref{def:jokers}). For any $q\in\reach$, we must have $q \in (\pre)^k(\jokerg)$ for some $k \in \mathbb{N}$, as $q$ can reach $R$ in a finite number of steps, because $Q$ is finite. Consequently, we have $q \in \jokerg$ too.

From above $\subseteq$-relations the stated equalities follow trivially. \qed

\hypertarget{proofthm4}{
\jokerstratcostminimal*
}

\noindent
\textit{Proof \autoref{thm:jokerstratwinning}.}

\noindent
(a) Let $q \in Q$ be a state and let $\sigma \in \Sigma_1(q)$ be a Joker attractor strategy according to \autoref{def:jokerstrat}.

First suppose that $q \notin \jokerg$. Then $q \notin \reach$ by \autoref{thm:jokerwinreach}, so $q$ has no winning play. Then any strategy -- also $\sigma$ -- has $\cost(\sigma) = \infty$ by \autoref{def:costfunction}.
Consequently, $\sigma$ is cost-minimal.

Now suppose that $q \in \jokerg$. We prove that $\sigma$ is cost-minimal by induction on $k = \jrankg(q)$.

If $k = 0$, then $\sigma$ is a standard attractor strategy that does not spend any Joker actions, so $\cost(\sigma) = 0$. A Joker game has no actions with negative costs, so trivially, $\sigma$ is cost-minimal.

For $k > 0$, we have the induction hypothesis that any Joker strategy $\sigma' \in \Sigma_1(q')$ with $\jrankg(q') = m$ for $m < k$ is cost-minimal.

If $q \in \jstate$, then $\sigma$ uses a Joker action $(a,x,q'')$ with $(q,(a,x,q'')) \in w\jokerg$. By \autoref{def:witness} we have that $q'' \in \joker{m}$ for $m < k$, so $\sigma$ is cost-minimal when starting in $q''$. By the Joker attractor construction we know that there is no possibility to reach $q''$ (or any other state in $\joker{m}$) from $q$ for sure (i.e. against any Player 2 and 3), because $q$ is not in the controllable predecessor of $\joker{m}$. So Player 2 or 3 can prevent Player 1 from winning if she uses no Joker action, which would yield a path, and corresponding strategy, with cost $\infty$ . Hence, using one Joker action is the minimum cost for reaching a state in $\joker{m}$. Consequently, $\sigma$ is cost-minimal in this case.

If $q \notin \jstate$, then $\sigma$ is a standard attractor strategy attracting to a Joker state of $\joker{k}$. These standard attractor actions have cost 0. Upon reaching a Joker state, the same reasoning as above applies: we need to spend one Joker action. Hence, $
\sigma$ is also cost-minimal in this case.

\noindent
(b) See \autoref{fig:costminnojokerstrat}. \qed

\hypertarget{proofthm5}{
 \jokerdetermined*
}
\noindent
 \textit{Proof \autoref{thm:jokerdetermined}.} This follows from \autoref{thm:jokerwinreach}:
 \begin{itemize}
     \item In any state $q \in \jokerg$ we have a winning Player 1 strategy.
     \item From any $q \notin \jokerg$, no state of $R$ can be reached, so Player 2 wins with any strategy. \qed
 \end{itemize}

\hypertarget{proofthm6}{
\admnonjoker*
}

\noindent
\textit{Proof \autoref{thm:admnonjoker}.}\\
See \autoref{fig:admnonjoker}. \qed

\hypertarget{proofthm7}{
\jokeradm*
}

\noindent
\textit{Proof \autoref{thm:jokeradm}.}\\
(1) See \autoref{fig:memstrat}.

\noindent
(2) Global cost-minimal strategies will, if possible, only use actions such that Player 2 and 3 cannot prevent Player 1 from winning, so we only need to consider states where Player 1 is not in full control: the Joker states. (If there are no Joker states, then any Joker-inspired cost-minimal strategy is winning.) We also note that Player 1 strategies that are not cost-minimal will not be able to do better, since at least the cost-minimal number of Joker states need to be passed by a play to a goal.
From any Joker state $q$, the game continues to a state $q'$, where either three of the following conditions hold:
\begin{enumerate}
    \item $q' \notin \reach$, i.e. Player 1 can never win the game any more.
    \item $\jrankg(q) > \jrankg(q')$, i.e. Player 1 got help from Player 2 and/or 3 to progress towards the goal.
    \item $q'\in \jokerg$ and $\jrankg(q) <= \jrankg(q')$, i.e. Player 1 is prevented to make progress.
\end{enumerate}
Note that, for any Player 1 action, there is a Player 2 and a Player 3 action that forces the game to a state of type (1) or (3), because otherwise $q$ would not have been a Joker state. Joker-inspired global cost-minimal strategies will only propose Player 1 actions that also have a Player 2 and a Player 3 action such that the game moves to a state of type (2). Because the Player 2 and 3 strategies are positional, their choice for the type of the next state, is fixed, given a Player 1 action. 

Hence, we draw the following conclusion (C): if two Player 1 strategies choose the same action, the next state of the game is the same. If two Player 1 strategies choose a different Player 1 action, then, depending on the game topology (i.e. the moves from state $q$), Player 2 and 3 either need to continue to the same type of state for any Player 1 action, or there are Player 2 and 3 strategies to make one Player 1 strategy continue to a type (2) state and the other to a (1) or (3) state, \emph{and vice versa} (because $q$ is a Joker state, so Player 1 cannot force the game to a type (2) state for any enabled Player 1 action).

Consequently, we have a subset of Joker states $Q_1 \subseteq \jstate$, where Player 2 and 3 can force the game to a type (1) state, where Player 1 cannot win (i.e. cannot ever reach a goal state). Furthermore we have a subset of Joker states $Q_3 \subseteq \jstate$, where Player 2 and 3 can only force the game to a type (3) state, but, after a finite number of moves, Player 2 and 3 can force the game to visit a state from $Q_1$ for any Player 1 strategy that attempts to reach a goal state (i.e. our global cost-optimal strategies). 

For the remaining set of Joker states $Q_3' = \jstate\setminus(Q_1 \cup Q_3)$, we show that Player 2 and 3 can force the game into a cycle, for any Player 1 strategy where Player 2 and 3 cannot force the game to a state of $Q_1 \cup Q_3$. Concretely, in any state $q \in Q_3'$, we have the following observation: Player 2 and 3 must be able to force the game to the same or another state $q' \in Q_3'$. 
Hence, we can think of this as a graph problem, where we have a finite set of nodes $Q_3'$, and each node needs to have an outgoing directed edge to a node of $Q_3'$. The analysis of this problem is as follows. When trying to add an edge $e$ for each node $n$, one by one, without making a cycle, one has to connect the edge from $n$ to a node that does not have and outgoing edge yet, because otherwise we create a cycle via $e$. Because the set $Q_3'$ is finite, the last added edge needs to go to a node that already has an outgoing edge. Consequently this creates a cycle anyway. Hence, also in Joker states $Q_3'$ Player 2 and 3 have a strategy such that Player 1 does not ever reach a goal state, as the game is kept in a cycle forever.

Consequently, from any Joker state, Player 2 and 3 have a strategy against any Player 1 strategy, such that a goal state is never reached. 
Also, Player 2 and 3 have a strategy that allows any global cost-minimal Player 1 strategy to reach a goal state.
Because the Player 2 and 3 strategies need to be positional, such a pair fixes the choice of Player 2 action and next state, for a given Player 1 action and current state. Hence, only global cost-minimal Player 1 strategies that choose different actions in Joker states, or steer the game to different Joker states via non-Joker states, may have different winning sets of Player 2 and 3 strategies.

This leads to the generalized conclusion C: two global cost-minimal Player 1 strategies either have the same pairs of winning Player 2 and 3 strategies, or for each of the two Player 1 strategies there is a pair of winning Player 2 and 3 strategies the other strategy cannot win from. Hence, each of the two global cost-minimal Player 1 strategies does not dominate the other strategy. Strategies that are not global cost-minimal will not be able to dominate either, as they also will have to pass a Joker state towards a goal state. Consequently, global cost-optimal strategies are admissible. \qed

\hypertarget{proofthm8}{
\shortcostminimal*
}

\noindent
\textit{Proof \autoref{thm:shortcostminimal}.}\\
(a) 

The following observation provides the foundation for below proof: 
When a Joker state $q$ is added to a set $\dattrk{k}$ of the Joker distance attractor (\autoref{def:dattr}) by the distance predecessor $\dpre$ (\autoref{def:dpre}), the definition of $\dpre$ requires the following when adding a state $q$:
\begin{itemize}
    \item When $\jrankg(q) = \jrankg(q') + 1$, $q$ must have been added for some Joker move from $q$ to $q'$, since this part of the definition (i.e. $\exists a \in \Gamma_1(q), \exists x \in \Gamma_2(q), \exists q' \in \moves(q,a,x):  q' \in Q'$) is exactly the predecessor $\pre$ operation  of \autoref{def:predecessors}.
    \item When $\jrankg(q) = \jrankg(q')$, $q$ is added for some controlled move from $q$ to $q'$, since this part of the definition (i.e. $\exists a \in \Gamma_1(q), \forall x \in \Gamma_2(q): \moves(q,a,x) \subseteq Q'$) is exactly the controlled predecessor of \autoref{def:predecessors}. So , the rank of the  states via the controlled predecessor remain the same, and no Joker action is used.
\end{itemize}

Now we do a proof by induction to show that Joker distance strategies are cost-minimal. 
Let $\sigma \in \Sigma(q^0)$ be a Joker distance strategy for Joker game $\jokergame$ with initial state $q^0 \in \jokerg$.
Let $\jrankg(q^0) = n$ for some $n \in \mathbb{N}$.
\begin{itemize}
    \item First suppose $n=0$. 
    Because the Joker attractor added $q^0$ via controlled predecessor operations only, this must be the case for the Joker distance predecessor too. Hence, the resulting strategy will enforce a play with states 
    $q^0, q',...,r$, where $r \in \finally$, all states have rank 0, and 0 Joker actions are used. Then trivially $\sigma$ is cost-minimal.
\item Now suppose that $n > 0$. From the induction hypothesis we obtain that for any state $q'$ with $\jrankg(q') = n-1$, the Joker distance strategy $\sigma$ will use $\jrankg(q')$ Jokers for any play starting from $q'$, i.e., against any opponent. 
    There must be a state $q'$ with $\jrankg(q') = n-1$ such that there is a play from $q$ to $q'$, because $q \in \joker{n}$ and $q'\in \joker{n-1}$. In particular, we know from the Joker attractor construction that we have a finite play $\pi$ from $q$ to $q'$ with states $q,...,q'',q_j,q''',...,q'''',q'$, such that $q''',...,q''''$ are obtained via the controlled predecessor, then state $q_j$ via the predecessor, i.e. via a Joker move, and states $q,..q.,q''$ via the controlled predecessor again. Because the Joker distance predecessor adds states also via the predecessor or controlled predecessor, while ensuring a rank increase of 1 only occurs when applying the predecessor, and otherwise remains equal, $q$ must also have be added to the Joker distance attractor at some iteration, though the intermediate states added can be different than those in $\pi$ (and hence the corresponding strategy differs). The consequence is that the Joker distance  attractor will have witnesses for moves from $q$ to $q'$ such that only one Joker action is used. By the induction hypothesis we then obtain that the Joker distance strategy uses $n$ Jokers from $q^0$.
\end{itemize}
Since Joker strategies are cost-minimal (\autoref{thm:jokerstratwinning}), and since have proven that Joker distance strategies use the same number of Joker actions, we obtain that Joker distance strategies are cost-minimal.\\

\noindent
(b)
We observe that the Joker distance predecessor selects all states that can be added via both the predecessor and the controlled predecessor. The Joker attractor either uses the controlled predecessor or the predecessor to add states. Consequently, the witnessed Joker distance attractor considers all and more winning plays in the Joker game than the witnessed Joker attractor. Moreover, the Joker distance attractor minimizes the distance of the selected moves because of its attractor structure, where its rank is the distance. Therefore a Joker distance strategy uses the same or less moves than a Joker strategy for any play. \qed

\hypertarget{proofthm9}{
\probjokercostless*
}

\noindent
\textit{Proof \autoref{thm:probjokercostless}.}

\noindent
This follows from a result of \cite{bordais_et_al22}. They define a set of states $Sure(R)$ that have a sure winning strategy (win against any opponent), i.e. this corresponds to $\attrf$ in our notation. They furthermore define a set of states $Almost(R)$ that have an almost sure winning strategy (win against any opponent with probability 1), i.e. this corresponds to $\pattrf$ in our notation. The result of \cite{bordais_et_al22} is then that $Sure(R) \subseteq Almost(R)$. Hence, the probabilistic attractor consists of the same or more states than the standard attractor. The result is that states that are Joker states according to the Joker attractor may be included in the probabilistic attractor. The probabilistic Joker attractor then finds that the initial state has a lower Joker rank, than with the Joker attractor, such that the randomized winning strategies $\sigma$ in the Joker game use less Jokers than Joker attractor strategies $\sigma'$. \autoref{fig:pennyext} shows that games exist for which this occurs. In the other cases (since  $Sure(R) \subseteq Almost(R)$), the probabilistic attractor and standard attractor yield the same set of states.   Hence, we have for randomized Joker strategies $\sigma$ and Joker attractor strategies $\sigma'$ that $\cost(\sigma) \le \cost(\sigma')$. \qed

\hypertarget{proofthm10}{
\probnrjokermoves*
}

\noindent
\textit{Proof \autoref{thm:probnrjokermoves}.}

\noindent
(1) We can follow the original proof of \autoref{thm:nrjokermoves}(1), and only make the following two changes to use it in the setting of randomized strategies:
\begin{itemize}
    \item In case $q' \in \attrf$, we have that $\sigma_1^J$ is a \emph{probablistic} attractor strategy from $q$ that is \emph{almost} sure winning, i.e. reach R \emph{with probability 1}, without using any Joker actions.
    \item In case $ q'\in \joker{k}\setminus\jstate$, we similarly have a \emph{probablistic} attractor strategy now, that \emph{almost} surely reaches one of the Joker states without using any Joker actions.
\end{itemize}

\noindent
(2) follows from \autoref{thm:probjokerstratwinning} as in the original proof of \autoref{thm:nrjokermoves}(2). \qed

\hypertarget{proofthm11}{
\probjokerwinreach*
}

\noindent
\textit{Proof \autoref{thm:probjokerwinreach}.}
We follow the proof of \autoref{thm:jokerwinreach} with the following changes:
\begin{itemize}
    \item In the proof of case $\reach \subseteq \winrop(\jokergame,\finally)$ we adapt the strategy $\sigma$ to play the proposed Joker actions with probability 1.
    \item In the proof of case $\winrop(\jokergame,\finally) \subseteq \reach$, we use for fact (1) that there is an almost sure winning strategy, and in fact (3) we use a Joker action that is played with probability 1 .
    \item The proof of case $\pjokerg \subseteq \winrop(\jokergame,\finally)$ uses lemma (W) of \autoref{thm:nrjokermoves}(1) as described for this case in \autoref{thm:jokerwinreach}. But in lemma (W) we substitute $\attrop$ by $\pattrop$, $\jokerg$ by $p\jokerg$, and $\jrankg$ by $p\jrankg$.
    \item In the proof of case $\reach \subseteq \pjokerg$, we need no changes, as the definition of $\pjokerg$ uses $\pre$ also when $\pattrf$ is used instead of $\attrf$. \qed
\end{itemize}

\hypertarget{proofthm12}{
\probjokerstratwinning*
}

\noindent
\textit{Proof \autoref{thm:probjokerstratwinning}.} We follow the proof of \autoref{thm:jokerstratwinning}, but use the probabilistic definitions at all places in this proof. Hence we adapt the proof as follows:
\begin{itemize}
    \item We define $\sigma \in \Sigma_1^r(q)$ as a randomized Joker strategy.
    \item In the first case we suppose that $q \notin \pjokerg$, and use \autoref{thm:probjokerwinreach} instead of \autoref{thm:jokerwinreach}.
    \item In the second case we take $q \in \pjokerg$ and prove that $\sigma$ is cost-minimal by induction on $k = \pjrankg(q)$.
    \begin{itemize}
        \item In case $k=0$ we use a randomized Joker strategy (\autoref{def:pattr}) instead of a standard attractor strategy, and similarly use no Joker actions, so $\sigma$ is cost-minimal.
        \item In case $k > 0$ we have the induction hypothesis that any $\sigma' \in \Sigma_1^r(q')$ with $\pjrankg(q') = m$ for $m < k$ is cost-minimal.
        \begin{itemize}
            \item If $q \in \pjstate$, then $\sigma$ uses a Joker action $(a,x,q'')$ with $(q,(a,x,q'')) \in w\pjokerg$. By \autoref{def:pjokers} we have that $ q'' \in \pjoker{m}$, so $\sigma$ is cost-minimal when starting in $q''$. By the probabilistic Joker attractor construction we know that there is no possibility to reach $ q''$ from $q$ with probability 1, because $q$ has been excluded by $\sattr_1$ at some point. For same reasons as in the proof of \autoref{thm:jokerstratwinning} it is minimum cost to spend 1 Joker action, so $\sigma$ is cost-minimal in this case.
            \item If $q \notin \pjstate$ then $\sigma$ is a randomized Joker strategy to a Joker state of $\pjoker{k}$, using 0 cost actions only. Upon reaching a Joker state we follow the above reasoning. Hence $\sigma$ is also cost-minimal in this case. \qed
        \end{itemize}
    \end{itemize}
\end{itemize}

\hypertarget{proofthm13}{
\probjokerdetermined*
}

\noindent
\textit{Proof \autoref{thm:probjokerdetermined}.}
Similar to the proof of \autoref{thm:jokerdetermined} this follows from \autoref{thm:probjokerwinreach}. \qed

\hypertarget{proofthm14}{
 \jokernotrandom*
 }

\noindent
\textit{Proof \autoref{thm:jokerrandom}.}
Let $\sigma_1 \in \Sigma_1^r(q^0)$ be an almost sure winning randomized strategy. \\
1) Then there is a winning play in $\Outc(\sigma_1)$. As in the proof of \autoref{thm:jokerwinreach}, we can use this play for constructing a strategy that plays the Joker actions matching the play.\\
2)
Because we use the probabilistic Joker attractor (\autoref{def:pattr}), in each probabilistic Joker attractor set, we can attract with probability 1 to a Joker state \cite{bordais_et_al22}. In a Joker state $q$ we choose Joker action $(a,x,q')$ such that $\pjrankg(q) > \pjrankg(q')$ with probability 1. By choosing such a move in a Joker state, the game moves to a state with a smaller Joker rank with probability 1. Decreasing the Joker rank in each Joker state leads to reaching a goal state. Because a next Joker state is reached with probability 1 each time, the overall probability of the resulting strategy is 1 too, so it is almost sure winning.\\
3) See the proof of 2). \qed\\

\hypertarget{proofthm15}{
\probjokerstrat*
}

\noindent
\textit{Proof \autoref{thm:probjokerstrat}.}
This is exactly what we use in the proof of \autoref{thm:jokerrandom}(2). \qed

\end{document}